\documentclass[prb,twocolumn,showpacs]{revtex4-1}
\usepackage[dvips]{graphicx}
\usepackage{epsfig}
\usepackage{color}
\usepackage[normalem]{ulem}
\usepackage{amsmath,amsfonts,amssymb,bm}
\setcitestyle{numbers,square}

\newcommand{\beginsupplement}{%
        \setcounter{table}{0}
        \renewcommand{\thetable}{S\arabic{table}}%
        \setcounter{figure}{0}
        \renewcommand{\thefigure}{S\arabic{figure}}%
        \setcounter{equation}{0}
        \renewcommand{\theequation}{S\arabic{equation}}%
        \setcounter{subsection}{0}
        \renewcommand{\thesubsection}{S\arabic{subsection}}
     }

\begin{document}
\title{Analytical theory of many-body localization in the presence of periodic drive}
\author{Alexander L. Burin}
\affiliation{Department of Chemistry, Tulane University, New
Orleans, LA 70118, USA}

\date{\today}
\begin{abstract}
Many-body localization transition in a periodically driven quantum system  is investigated using a solution of a matching Bethe lattice problem for  Floquet states of a quantum random energy model with a generalization to more realistic settings. It turns out  that an external periodic field can both suppress and enhance localization depending on field amplitude and frequency which leads to three distinguishable regimes of field enhanced, controlled and suppressed delocalization. The results can be verified experimentally in systems of cold atoms and/or interacting spin defects in semiconductors. 
\end{abstract}

\maketitle


Many body localization (MBL) suggests the lack of thermalization in interacting quantum systems if a many-body (or non-linear) interaction is sufficiently weak  \cite{KAM1}.  This phenomenon  is important in quantum informatics since delocaization and thermalization erase quantum memory \cite{Polkovnikov11RevModPhys}; it  has been considered in many physical systems including interacting localized electrons  \cite{FleishmanAnderson80,Altshuler82,Silvan94Exp,AltshullerGefen97,Gornyi05,Basko06,
ab16GutmanMirlin,Ovadyahu15MBL,Ovadia15},
quantum defects in a $^{4}$He crystal \cite{KaganMaksimov85}, tunneling two level systems (TLSs) in amorphous solids \cite{ab98book,ab01acFprl}, 
anharmonic vibrations in polyatomic molecules  \cite{Stewart83,LoganWolynes90,15LeitnerReview} and  spin excitations in semiconductors and chains of trapped ions \cite{Bloch16PerDriveExp,
Yao14MBLLongRange,Andraschko14,Monro16TimePerCryst,LukinDiamond16}.  

MBL can take place in a non-stationary regime under the  presence of a periodic drive \cite{Russomanno12,Lazarides14PRE,Alessio14,Abanin15TimeDep,
Lazarides15TDep,Demler16,Abanin16TimeDepRig,Pollmann16EnhNum}. This regime is fundamentally and practically significant in the specific case of periodic sinusoidal drives because such drives are widely used experimentally to determine system response functions \cite{Bukov15RevFloquetEng,Huse15LowFrCond,
Enss94LowFreqMeas,ab01acFprl} and to control a system quantum state \cite{ab13LZTh,ab16Bahman,ab16TLSlaserExp}. It has been recently suggested that a  periodic drive can create novel phases of matter  with broken time-translational symmetry  \cite{Khemani16,Yao16Floq,Monro16TimePerCryst}.  

Currently it has been demonstrated both numerically  and analytically that external drive can substantially suppress localization if the field amplitude $\hbar\delta$ exceeds its quantization energy $\hbar\omega$  \cite{Abanin15TimeDep,Demler16,Lazarides15TDep,LazaridesEqFl,Pollmann16PerDrive,Bloch16PerDriveExp}. 
Yet an MBL transition in periodically driven systems is not fully understood because of the obvious limitation of numerical studies to relatively small systems and the lack of the analytical theory (notice, however, the remarkable progress attained in Ref. \cite{Abanin16Period}). 

Consequently the development of the analytical theory would be beneficial for characterizing an MBL transition in the presence of a periodic drive. Here the step towards such development is proposed for a random energy quantum spin glass model \cite{Derrida80,Laumann14} with a sinusoidal periodic drive (the generalization to a binary discontinuous drive \cite{Lazarides15TDep,Huse16Floq} is addressed qualitatively in the end of the present work). The  localization transition is described in this model (see Eq. (\ref{eq:ANSMain0}) and Figs. \ref{fig:GamGamc}, \ref{fig:GamcW} and Figs. S5, S6 in Supplementary Materials) in a wide range of field amplitudes $\hbar\delta$ and frequencies $\omega$. Three distinguishable field effects on delocalization are revealed in  different parameter domain including field enhanced, field controlled and field suppressed delocalization. These results can be extended to more realistic settings with possible modifications of logarithmic factors in Eq. (\ref{eq:ANSMain0})  (see discussion after Eq. (\ref{eq:ANSMain0})).


Consider a random energy spin glass model in the transverse field \cite{Laumann14} driven by  a sinusoidal periodic longitudinal field that can be described by the Hamiltonian
\begin{eqnarray}
\widehat{H}=\Phi(\{\widehat{\sigma}_{i}^{z}\})-\Gamma\sum_{i}^{N}\widehat{\sigma}_{i}^{x}-\frac{\hbar\delta \cos(\omega t)}{2}\sum_{i}^{N}\widehat{\sigma}_{i}^{z},
\nonumber\\
P(\Phi)=\frac{1}{\sqrt{2\pi}W}e^{-\frac{\Phi^2}{2W^2}}.  
\label{eq:HSG}
\end{eqnarray}
The first term in the Hamiltonian represents a random term diagonal in the basis of product states with fixed spin projections to the $z$-axis and with sequence dependent random energies $\Phi \sim W$ 
uncorrelated in different states and characterized by a Gaussian distribution $P(\Phi)$. Second and third terms stand for constant transverse and sinusoidal, periodic (with a frequency $\omega$) longitudinal fields with amplitudes  $\Gamma$ and $\hbar\delta$, respectively  \cite{Laumann14,Demler16,Pollmann16PerDrive}. The typical random energy $W$ scales with the number of spins as $W = w\sqrt{N}$ , where $w$ is a characteristic diagonal energy per spin ($w=1/\sqrt{2}$ in Ref. \cite{Laumann14}).

\begin{figure}[h!]
\centering
\includegraphics[width=\columnwidth]{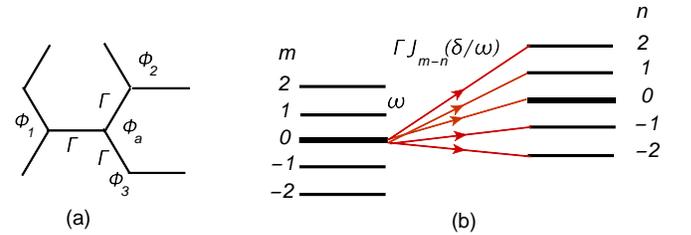}
\caption{\small (a) A Bethe lattice model for connectivity $K=3$. (b) Spin flip transitions involving different Floquet states.}
\label{fig:Interf}
\end{figure}

Consider the localization problem for states with energies close to zero corresponding to an infinite temperature. In the absence of a drive the problem, Eq. (\ref{eq:HSG}), can be matched with the localization problem on the Bethe lattice \cite{Laumann14,ab16preprintSG}. Each Bethe lattice site $a$ is determined by an $N$ spin projection sequence $a=\{\sigma_{i}^{z}=\sigma_{ia}\}$ representing an Ising model eigenstates (see Fig. \ref{fig:Interf}.a). Each site is connected by the transverse field $\Gamma$ to $K$ other sites $a_{k}$ ($k=1,...N$) different from the given state by a single spin flip ($\sigma_{k}^{z}=-\sigma_{ka}$) where $K=N$ is a Bethe lattice coordination number.   The interference between different paths (spin flips taken in a different order) absent in an exact Bethe lattice problem and existing in a spin glass model can be approximately neglected for a large number of spins $K \gg 1$ \cite{Laumann14,ab16preprintSG}. In the absence of a periodic drive an accurate solution for the localization threshold at zero energy  expressed in terms of a minimum transverse field $\Gamma_{c0}$ needed for delocalization of states with energy $E=0$ takes the form (see Refs. \cite{AbouChacra73,Laumann14,ab16preprintSG,Aizenman11} and Supplementary Materials, Sec. S1, remember that the coordination number $K$ is equal to the number of spins $N$ in a random energy model)
\begin{eqnarray}
\Gamma_{c0} \approx  \frac{1}{4P(0)K\ln(K)}.   
\label{eq:BetheThr}
\end{eqnarray}
The localization threshold reaches the minimum at a zero energy \cite{Laumann14} and this minimum, $\Gamma_{c0}$, will be used below as a reference point.    

In a periodically driven system, Eq. (\ref{eq:HSG}), quasi-energies and Floquet states should be used  instead of eigenenergies and eigenstates. For a single site $a$ a wavefunction amplitude $c_{a}$ for a Floquet state characterized by a quasi-energy $\epsilon$ ($-\hbar\omega/2<\epsilon<\hbar\omega/2$) can be expressed  within the  rotating frames with a time-dependent rotation frequency \cite{ab01acFprl,Deng15BesselFloq,Bukov15RevFloquetEng} as $c_{a}=e^{-i\epsilon t/\hbar+\frac{i\delta \sigma_{a}\sin(\omega t)}{2\omega}}\sum_{n}c_{a n}e^{i\omega nt}$, where $\sigma_{a}=\sum_{i=1}^{N}\sigma_{a i}$ is the projection of the total spin in the product state $a$ to the $z$-axis.
In this representation one can express the Shr\"odinger equation for the wavefunction $c_{an}$ as \cite{ab01acFprl} 
\begin{eqnarray}
(\epsilon-m\hbar\omega-\Phi_{a})c_{a m} = \Gamma\sum_{k,n}J_{m-n}(\delta\sigma_{a k}/\omega)c_{k n}.  
\label{eq:FloqShr}
\end{eqnarray}
For a sufficiently small frequency  $\omega \ll W/\hbar$ Floquet states with different quasi-energies corresponds to close random energies  and their localization and delocalization occur approximately simultaneously so one can characterize the MBL transition using the states with $\epsilon=0$. At larger frequencies localization threshold depends on the quasi-energy in a random energy model similarly to Ref. \cite{Laumann14}.  Since this dependence does not show up in more realistic settings \cite{ab16preprintSG} its consideration for a random energy model is moved to Supplementary Materials, Secs. S1, S3. 


The localization threshold depends on the parameters $\delta$, $\omega$, $W$ and $K$. In the case of a small driving field, $\delta \ll \omega$, one has $J_{m-n}(\delta\sigma_{a k}/\omega) \approx \delta_{mn}$ where $\delta_{mn}$ is a Kronecker delta symbol \cite{GradshteynRyzhik07,Abanin15TimeDep,Demler16}. Then Eq. (\ref{eq:FloqShr}) becomes approximately insensitive to the drive and the localization transition at a zero quasi-energy is determined  by $\Gamma_{c0}$, Eq. (\ref{eq:BetheThr}).  

In the remaining case $\delta > \omega$ one can qualitatively approximate Bessel functions in Eq. (\ref{eq:FloqShr}) as \cite{GradshteynRyzhik07,ab01acFprl}
\begin{eqnarray}
J_{n}(\delta\sigma_{a k}/\omega) \approx 
\begin{cases}
    \sqrt{\omega/\delta}, & \text{if $n< \delta/\omega$},\\
    0, & \text{otherwise}.
  \end{cases}, ~ \delta > \omega.   
\label{eq:Limits00}
\end{eqnarray}
Using this approximation for the Bessel functions determining Floquet states interaction one can qualitatively describe the localization transition and determine the localization threshold with the accuracy to numerical factors of order of $1$. This qualitative derivation is given below while the more accurate analysis employing the Green function method is reported in Supplementary Materials (Sec. S2). Asymptotic analytical dependencies for different parameter domains are given in Eq. (\ref{eq:ANSMain0}) and in Fig. \ref{fig:GamcW} and they are consistent with the more accurate numerical solution derived in Supplementary materials (Eq.  S31). These numerical solutions are shown in Figs. \ref{fig:GamGamc}, \ref{fig:GamcW}. 

The localization threshold behaves differently in different parameter domains including intermediate drive ($W/K < \hbar\delta < W$), strong drive at low frequency ($\hbar\omega < W < \hbar\delta$), strong drive at high frequency ($W < \hbar\omega < \hbar\delta$) and weak drive ($\hbar\omega < \hbar\delta < W/K$). At an intermediate driving field,  $W/K < \hbar\delta < W$, each coupling $\Gamma$ of two states splits into $n_{F} = \delta/\omega$ couplings $\Gamma J_{m}(\delta/\omega) \sim \Gamma/\sqrt{n_{F}}$ ($|m|<n_{F}$) of a given state to different Floquet states (see Fig. \ref{fig:Interf}. b and Eq. (\ref{eq:Limits00})). A resonant interaction can take place for each out of $n_{F}$ couplings under the condition $|\Phi_{k}-\Phi_{a} -m\hbar\omega| < \Gamma/\sqrt{n_{F}}$. If resonant interactions can be considered as independent the probability of resonance increases by the factor $\sqrt{n_{F}}$ compared to the system without periodic drive. Since delocalization takes place in the presence of an approximately one resonance per state \cite{AbouChacra73,Laumann14,ab16preprintSG,Shepelyansky98} one can expect the reduction of localization threshold as $\Gamma_{c} \sim \Gamma_{c0}/\sqrt{n_{F}} \sim \Gamma_{c0} \sqrt{\omega/\delta}$. Under this condition resonances can be indeed treated as independent since their energy splitting $\hbar\omega$ exceeds the coupling strength $\Gamma_{c}\sqrt{\omega/\delta}$. Indeed, using Eq. (\ref{eq:BetheThr}) and the initial assumption  $W/K < \hbar\delta$ one can see that $\Gamma_{c}\sqrt{\omega/\delta} \sim \Gamma_{c0}\omega/\delta < \hbar\omega (W/(K\hbar\delta)) \ll \hbar\omega$. The change in delocalization mechanism compared to the system without periodic drive suggests the modification of the logarithmic factor in the definition of the localization threshold  given by $\ln(K)^{-1}$ in Eq. (\ref{eq:BetheThr}). In the self-consistent theory of localization the argument of logarithm is given by the ratio of the maximum detuning from resonance ($W$ in the absence of an external drive) and the minimum detuning ($\Gamma^2/W$) needed to avoid a strong modification of the real part of self-energy \cite{AbouChacra73}. The logarithm of the ratio of upper and lower limits  yields $\ln(K)$ at $\Gamma \sim \Gamma_{c0}$. Here the maximum detuning is given by the field quantization energy $\hbar\omega$ since at larger detuning the other Floquet resonance will be dominating. The minimum detuning is determined by the reduced coupling strength (see Fig. \ref{fig:Interf}. b, Eq. (\ref{eq:Limits00})), as $\Gamma^2 \omega/(\delta W)$. These modified upper and lower integration limits determine the ratio $\ln(K)/\ln(\xi)$  in Eq. (\ref{eq:ANSMain0}) (second and third lines). The analysis of localization threshold in Supplementary Materials employing the Green function method (Sec. S2) determines the numerical factors as well, and these factors are included in Eq. (\ref{eq:ANSMain0}). 

Thus in this intermediate regime, $W/K < \hbar\delta < W$, the delocalization is always enhanced by the external drive so it can be referred as a {\it field enhanced delocalization} (FED). The reduction of the localization threshold in this regime compared to Eq. (\ref{eq:BetheThr}) is shown in Fig. \ref{fig:GamGamc} at large random energies $W$ and in Fig. \ref{fig:GamcW} (dashed lines) at small and intermediate  field amplitudes.

In the case of a large field, yet a small frequency, $\hbar\omega \ll W \ll \hbar\delta$, the number of delocalization channels gets fixed at $n_{1F} \sim W/\hbar\omega$ because the probability to find resonances at energies exceeding $W$ is exponentially small, Eq. (\ref{eq:HSG}). Yet the coupling strength is reduced by the factor $1/\sqrt{n_{F}}$ suggesting $\Gamma_{c} \sim \Gamma_{c0} n_{1F}/\sqrt{n_{F}} \sim \hbar\sqrt{\delta\omega}/K$. Thus the delocalization transition is almost insensitive to the characteristic system energy $W$ except for the weak logarithmic dependence, so this regime   can be referred as a {\it field controlled delocalization} (FCD). It corresponds to intermediate random energies in Fig. \ref{fig:GamGamc} or large amplitudes in Fig. \ref{fig:GamcW} (dotted lines there). The logarithmic dependence, Eq. (\ref{eq:ANSMain0}),  leads to a slow reduction in $\Gamma_{c}$ with increasing $W$.

In the case of a large quantization energy compared to a typical random energy, $\delta > \omega > W/\hbar$, a resonant interaction can only take place between states with identical Floquet indices $n$ in Eq. (\ref{eq:FloqShr}). Indeed, the probability of resonance with different Floquest state $P(\hbar\omega)\sim e^{-(\hbar\omega/W)^2/2}$, Eq. (\ref{eq:HSG}),  is exponentially small. This reduces the problem in Eq. (\ref{eq:FloqShr}) to the quantum random energy model without periodic drive \cite{Laumann14} and with a modified transverse field $\Gamma_{1}=\Gamma J_{0}(\delta/\omega)$ similarly to Ref. \cite{Das10}. Consequently the localization threshold increases by the factor  $|J_{0}(\delta/\omega)|^{-1}$, see Eq. (\ref{eq:ANSMain0}) and Figs. \ref{fig:GamGamc}, \ref{fig:GamcW}. This regime   can be referred as a {\it field suppressed delocalization} (FSD). The expression for the localization  threshold should be modified near Bessel function zeros, $J_{0}(\delta/\omega)=0$,  where the localization threshold formally approaches infinity. The upper constraint for the localization threshold is determined is Supplementary Materials (Sec. S3) and included into Eq. (\ref{eq:ANSMain0}), fourth line there. 


Finally consider the remaining case of a small driving field, $\hbar\delta \ll  W/K$, Eq. (\ref{eq:BetheThr}). The analysis below follows the earlier consideration \cite{Abanin16Period}  and leads to qualitatively similar behaviors with an advantage that the phenomenological parameters introduced there are determined here. 
The driving field is capable to stimulate a transition  between states with energies separated by the distance less than $\hbar\delta$ \cite{Abanin16Period} where a real level crossing by the periodic drive takes place. The minimum  energy splitting between the given state and the one of the adjacent states in Fig. \ref{fig:Interf}. a can be estimated as $W/K$, which exceeds   $\hbar\delta$ in the case of small fields.  Level crossing can take place with one of $K_{p}$ states in the $p^{th}$ coordination sphere, $K_{p} \approx K^{p}$ (it is assumed that $p \ll K$, see Sec. S4 in Supplementary Materials for detail), if the field amplitude $\sqrt{p}\hbar\delta$ exceeds a minimum energy splitting $W/K_{p}$ (cf. the estimate $e^{-p/\xi}$ in Ref. \cite{Abanin16Period} and the factor of $\sqrt{p}$ describes the average change in the total spin after $p$ flips). Consequently, $p \approx \ln(W/\hbar\delta)/\ln(K)$ for $1\ll p \ll K$.  

Assume that $\Gamma<\Gamma_{c0}$. Then in the absence of a periodic drive the probability of resonant interaction within the $p^{th}$ coordination sphere can be expressed as $(\Gamma/\Gamma_{c0})^{p}$. The drive effect on resonant interactions can be estimated assuming that  there is no level crossing by the periodic field in the previous $p-1$ spheres. Then the probability of resonance due to level crossings in the $p^{th}$ sphere will be increased compared to the stationary problem by the factor of  $\sqrt{\delta/\omega}$ similarly to the case of intermediate fields $\hbar\delta>W/K$, Eq. (\ref{eq:ANSMain0}).  Then one can estimate the number of resonances as $\sqrt{\delta/\omega}(\Gamma/\Gamma_{c0})^{p}$. Setting this number to be equal unity leads to the localization threshold estimate in Eq. (\ref{eq:ANSMain0}) corrected by the logarithmic factor $L_{*}$ derived in Supplementary Materials. Since in the present case a delocalization is also enhanced by the field this is another regime of a {\it field enhanced delocalization}. The consideration is applicable to the random energy model if the driving field amplitude exceeds the minimum level splitting, $W/2^{K} \ll \hbar\delta$, while at smaller field its effect on the localization is negligible since it cannot induce level crossings. 

\begin{figure}[h!]
\centering
\includegraphics[width=\columnwidth]{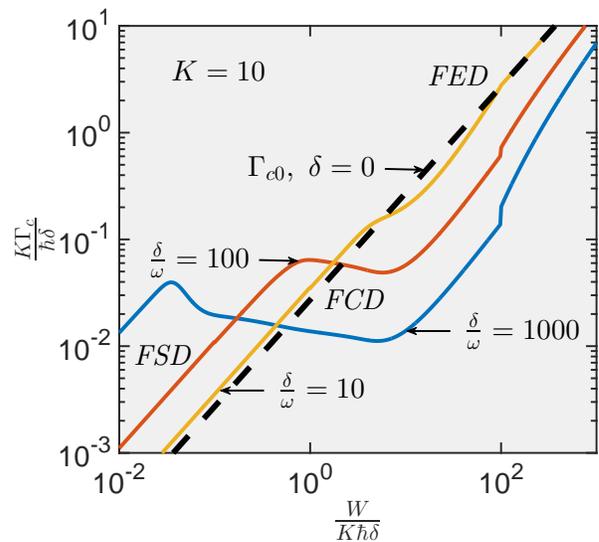}
\caption{\small Localization threshold vs. a typical spin-flip energy $W$. The dependencies are found solving numerically Eq. S31 in Supplementary Materials for $W/K<\delta$ and $\hbar\omega < W$ or otherwise using asymptotic solutions, Eq. (\ref{eq:ANSMain0})).} 
\label{fig:GamGamc}
\end{figure}

The results of the above considerations can be summarized as 
\begin{eqnarray}
\frac{\Gamma_{c}}{\Gamma_{c0}} 
\approx\begin{cases}
\left(\frac{\omega}{\delta}\right)^{\frac{\ln(K)}{2\ln(W/\hbar\delta)}}\frac{\ln(K)}{\ln(\xi_{1})}, ~ \omega < \delta < \frac{W}{\hbar K},\\
\frac{1}{\sqrt{\delta/\omega}+0.61}\frac{\ln(K)}{0.6\ln(\xi)},
   ~ \frac{W}{\hbar K}<\delta < \frac{W}{\hbar},\\
    \frac{\pi \hbar\sqrt{\delta\omega}}{2W}\frac{\ln(K)}{\ln(\xi)}, ~ \omega< \frac{W}{\hbar} < \delta,\\
\frac{1}{{\rm max} \left(|J_{0}(\delta/\omega)|, ~ \frac{6^{1/4}\sqrt{\pi}W}{\hbar\sqrt{2\omega\delta\ln(K)}}\right)}, 
~ \frac{W}{\hbar} < \omega, 
  \end{cases}  
  \nonumber\\
\xi=\frac{K^2\hbar\delta^2 W}{\omega(W^2+(\hbar\delta)^2)}, ~  \xi_{1}=\frac{K}{\sqrt{\left(\frac{\omega}{\delta}\right)^{\frac{\ln(K)}{2\ln(W/\hbar\delta)}}+\frac{\hbar\delta}{W}}}.  
\label{eq:ANSMain0}
\end{eqnarray}
These asymptotic behaviors are consistent with more accurate numerical solutions  obtained in Supplementary Materials and shown in Figs. \ref{fig:GamGamc}, \ref{fig:GamcW} (see also Figs. S5, S6 in Supplementary Materials). Strong oscillations of the localization threshold at high frequency are due to the Bessel function quasi-periodic behavior with a passage through zero. {\it These order of magnitude oscillations can be used to switch the system between localized and delocalized regime using a small change in an external drive.}

\begin{figure}[h!]
\centering
\includegraphics[width=\columnwidth]{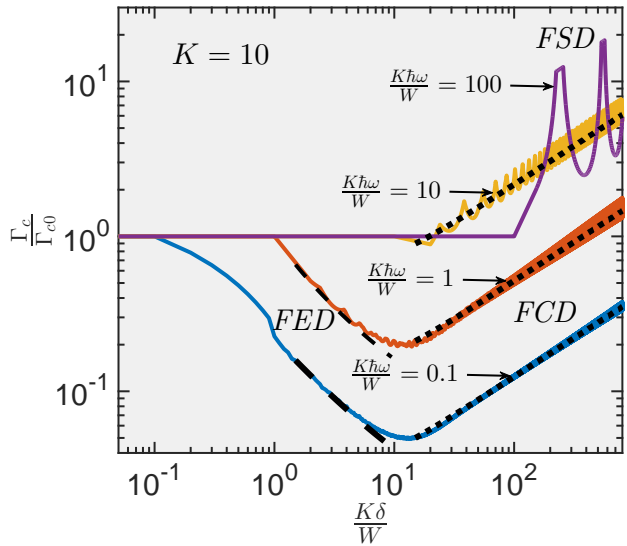}
\caption{\small Change in a localization threshold due to a periodic drive vs. the drive amplitude. The dependencies shown by solid lines were found similarly to those in Fig. \ref{fig:GamGamc}. The dashed and dotted lines show asymptotic behaviors for field enhanced (FED) and  field controlled (FCD) delocalization regimes (second and third lines in Eq. (\ref{eq:ANSMain0})), while a field suppressed regime (FSD) (fourth line) takes place at the highest frequency.} 
\label{fig:GamcW}
\end{figure}

The random energy model is not very realistic because the interaction there is of an infinite range and therefore localization threshold in Eqs. (\ref{eq:BetheThr}), (\ref{eq:ANSMain0}) vanishes in a thermodynamic limit $N \rightarrow \infty$. Can the results be extended to real systems with finite-range interactions?  

Consider parameters determining localization threshold, Eq. (\ref{eq:BetheThr}).  They include the characteristic energy difference of two neighboring states in the Bethe lattice $W \sim 1/P(0) \sim \sqrt{N}$ and the effective coordination number $K$ equal to the number of spins $N$ approaching infinity in a thermodynamic limit.  

In a more realistic model with a  finite-range interaction and delocalization still determined by the transverse field $\Gamma$  the energy difference of two ``neighboring" states is given by a spin flip energy $W$, which remains finite in a thermodynamic limit $N\rightarrow \infty$ \cite{ab16preprintSG}. The resonance probability is still determined by the distribution function $p(0) \sim 1/W$ of these flip energies at a zero energy \cite{Ros15,ab16preprintSG}. In the case of a finite-range interaction the coordination number $K$ of a matching Bethe lattice problem is also expected to be finite. It can be determined by a single particle  localization length or interaction radius $l$ \cite{AltshullerGefen97,Gornyi05,Basko06,ab90Kontor,abGorniyMirlinDot}. For instance in a $d$ dimensional system with an average distance $a$ between single particle localized states  one would expect $K \sim (l/a)^{d}$ while $K \sim 1$ for a nearest neighbor interaction \cite{Pollmann16PerDrive}. If $K \gg 1$ ($l \gg a$) then the problem still matches the Bethe lattice localization problem \cite{AltshullerGefen97,Gornyi05,Basko06} where the localization threshold, $\Gamma_{c0} \sim W/(K\ln(K))$, remains finite in the thermodynamic limit.  

Similarly to a random energy model the  effect of a periodic sinusoidal drive on a delocalization can be described in terms of the increased coordination number and reduced coupling strength. For instance, one would expect $K\rightarrow K_{F}\sim K\delta/\omega$ and $\Gamma \rightarrow \Gamma_{F}\approx \Gamma \sqrt{\omega/\delta}$ in the case of intermediate fields, $W/K < \hbar\delta < W$, Eq. (\ref{eq:Limits00}), and consequently $\Gamma_{c} \sim \Gamma_{c0}\sqrt{\omega/\delta}$ ($\omega \ll \delta$). Accordingly,   Eq. (\ref{eq:ANSMain0}) should remain valid at least qualitatively in systems with short-range interactions if one replaces $\Gamma_{c0}$ with the localization threshold for the related problem without drive. 

The case of a periodic binary drive with a discontinuous longitudinal field equal to $\pm \hbar\delta/2$ in the first and second halves of each period, respectively,  can be treated similarly to the sinusoidal drive (see Supplementary Materials, Sec. S6). One can also specify three delocalization regimes there in the same parameter  domains as for a sinusoidal drive with modified localization threshold behaviors (Eq. S51).

In summary, the external drive influence on MBL transition has been investigated and three different regimes of the field enhanced, controlled and suppressed delocalization have been revealed. Strong oscillations of localization threshold  at large field amplitudes can be used to control  the system state by small changes in the drive.  The results call for the experimental verification particularly using spin excitations in arrays of cold atoms where the perodic sinusoidal drive has been already  realized experimentally  \cite{Monro16TimePerCryst,Bloch16PerDriveExp}. 

The consideration of MBL transition has been restricted to infinite temperature states.  It is not clear \cite{Abanin15AoPSharpPert,Lazarides14PRE}  whether a periodic drive would heat these states to an infinite  temperature even  if infinite temperature states are delocalized. Energies of infinite temperature states differ from those at  a finite temperature by macroscopic energy  \cite{Laumann14} and it is not obvious whether the periodic drive can overcome the associated barrier. This problem is beyond the scope of the present work.

This work is partially supported by the National Science Foundation (CHE-1462075) and the Tulane University Carol Lavin Bernick Faculty Grant. Author acknowledges stimulating discussions with Frank Pollmann and Achilles Lazarides during his visit to the Max Planck Institute for Complex Systems (Dresden) and the Institute for supporting his visit.

\bibliography{MBL}

\beginsupplement

\begin{widetext}
\section{Supplementary Materials}

\subsection{Energy dependent localization threshold in a random energy model in the absence of an external drive}
\label{sec:NoDrive}

The localization threshold in a random energy model can be investigated using the Green function method \cite{AbouChacra73}. A single site Green function can be defined as $G_{a}(E)=<a|(E-\widehat{H})^{-1}|a>$ where states $|a>$ are the states with fixed spin projections to the $z$ axis and the Hamiltonian is taken without a time-dependent field 
\begin{eqnarray}
\widehat{H}=\Phi(\{\widehat{\sigma}_{i}^{z}\})-\Gamma\sum_{i}^{N}\widehat{\sigma}_{i}^{x}, ~
P(\Phi)=\frac{1}{\sqrt{2\pi}W}e^{-\frac{\Phi^2}{2W^2}}.  
\label{eq:HSGSM}
\end{eqnarray}

Diagonal Green functions can be expressed as $G_{a}(E)=(E-\Phi_{a}-\Sigma_{a})^{-1}$ where $\Sigma_{a}$ is a self-energy. For the large coordination number $K=N\gg 1$ imaginary parts of self-energies satisfy the self-consistent equations \cite{AbouChacra73}
\begin{eqnarray}
{\rm Im} \Sigma_{a}(E)=\Gamma^2\sum_{k=1}^{K}|G_{k}|^2 {\rm Im}\Sigma_{k}(E),  
\label{eq:SCGFSI}
\end{eqnarray} 
where the sum is taken over all neighboring sites $k$ obtained from the given site $a$ by a flip of the $k^{th}$ spin. 

The delocalization transition takes place at the critical transverse field $\Gamma_{c0}(E)$ where Eq. (\ref{eq:SCGFSI}) acquires a first non-zero solution. To estimate this field one can use the procedure of Ref. \cite{AbouChacra73}  as 
\begin{eqnarray}
\left<\exp(-t{\rm Im} \Sigma_{a}(E))\right>
=\left<\exp\left(-Kt\Gamma^2|G_{k}(E)|^2 {\rm Im}\Sigma_{k}(E)\right)\right>,
\label{eq:AbChSI}
\end{eqnarray}
where averaging is performed over all random energies $\Phi$ and the parameter $t$ is greater than $0$ and considered in the limit $t \rightarrow 0$. Using the anzatz  \cite{AbouChacra73,ab16preprintSG} 
$\left<\exp(-t{\rm Im} \Sigma_{a}(E))\right>\approx 1-\sqrt{t}F(E)$, one can obtain the self-consistent equation for amplitudes $F(E)$ in the form $F(E)=K\Gamma F(E)<|G_{k}(E)|>$ which determines the localization threshold as
\begin{eqnarray}
1=K\Gamma_{c0}(E) <|G_{k}(E)|>.
\label{eq:AbCh2SI}
\end{eqnarray}
To evaluate the right hand side in Eq. (\ref{eq:AbCh2SI})  one can approximate the Green function there by its zeroth order expression $G^{k}_{n} \approx 1/(\epsilon-\Phi_{k})$ except for the energy domain  $|E-\Phi_{k}| < \Gamma_{c0}(E)^2/W$, where the associated correction to the real part of the self-energy exceeds the characteristic energy $W$. This domaiun can be excluded with the logarithmic accuracy, which is justified in the case of a large coordination number $K \gg 1$. Then the average Green function absolute value can be estimated as 
\begin{eqnarray}
<|G_{k}(E)|> = \int_{D}\frac{P(\Phi)d\Phi}{|E-\Phi|}, 
\label{eq:AbCh3SI}
\end{eqnarray}
where the integration domain $D$ is determined by the inequality $\Gamma_{c0}(E)^2/W<|\Phi-E|$.

Using the definition of the distribution function $P(\Phi)=e^{-\Phi^2/(2W^2)}/(\sqrt{2\pi}W)$ and substituting the integration variable as $\Phi=Wx$ one can approximate the integral in Eq. (\ref{eq:AbCh3SI}) as 
\begin{eqnarray}
<|G_{k}(E)|> = \frac{1}{\sqrt{2\pi}W}\int_{h}^{\infty}\left(e^{-\frac{(\xi+x)^2}{2}}+e^{-\frac{(\xi-x)^2}{2}}\right)\frac{dx}{x}, ~ h=\frac{W^2}{\Gamma^2}, ~ \xi=\frac{|E|}{W}.  
\label{eq:AbCh4SI}
\end{eqnarray} 
This integral can be evaluated by parts as 
\begin{eqnarray}
<|G_{k}(E)|> = -\frac{\left(e^{-\frac{(\xi+h)^2}{2}}+e^{-\frac{(\xi-h)^2}{2}}\right)}{\sqrt{2\pi}W}\ln(h)+
\frac{1}{\sqrt{2\pi}W}\int_{h}^{\infty}\left(e^{-\frac{(\xi+x)^2}{2}}(\xi+x)+e^{-\frac{(\xi-x)^2}{2}}(x-\xi)\right)\ln(x)dx.  
\label{eq:AbCh5SI}
\end{eqnarray} 

According to the localization threshold  estimate, Eq. (\ref{eq:AbCh6SI}), one can always assume $\Gamma_{c0}(E) \ll W$ and $h \ll 1$. Then the first term in Eq. (\ref{eq:AbCh5SI})  can be approximated  by $4P(E)\ln(W/\Gamma)$ within the logarithmic accuracy. In the second term one can set the lowest integration limit approximately to $0$. Then the average absolute value of the Green function can be expressed as 
\begin{eqnarray}
<|G_{k}(E)|> = 4P(E)\ln(W/\Gamma)+\frac{f(E/W)}{W}; 
\nonumber\\
f(\xi)=\frac{1}{\sqrt{2\pi}}\int_{0}^{\infty}\left(e^{-\frac{(\xi+x)^2}{2}}(\xi+x)+e^{-\frac{(\xi-x)^2}{2}}(x-\xi)\right)\ln(x)dx.  
\label{eq:AbCh5SIA}
\end{eqnarray} 
The numerical evaluation of the function $f(\xi)$ shows (see Fig. \ref{fig:EnDep}) that it can be well approximated by $1/\xi$ in the case of $\xi \gg 1$. For $\xi \leq 1$ the first term in Eq. (\ref{eq:AbCh5SI}) dominates and the second term can be ignored. Then one can approximately represent  the average Green function in the form $4P(E)\ln(W/\Gamma)+\theta(|E|-W)/|E|$. The localization threshold can be defined as 
\begin{eqnarray}
\Gamma_{c0}(E)\approx \frac{1}{K(4P(E)\ln(W/\Gamma)+\theta(|E|-W)/|E|)}.
\label{eq:AbCh6SI}
\end{eqnarray}  
Here $\theta(x)$ is a Heaviside theta function. The behaviors $\Gamma_{c0}(0)\approx 1/(4KP(0)\ln(W/\Gamma))$ for zero energy and $\Gamma_{c0}(E)\approx |E|/K$ at large absolute value of energy  corresponding to a finite temperature (remember that $K=N$) are both consistent with Ref. \cite{Laumann14} with the accuracy to the numerical factor $e/2$. The difference $e/2$ between the prefactors in the definitions of localization thresholds  has the same origin as in two different estimates in Ref. \cite{AbouChacra73} where the integration constraint is included (Sec. 6 there) or ignored (Sec. 5 there). The integration constraint   is included in the present work since it gives a better estimate for the localization transition in the Bethe lattice \cite{AbouChacra73}. The addition of a similar constraint to Ref. \cite{Laumann14} will change the estimate for the localization threshold by the same factor so two considerations are technically equivalent.

\begin{figure}[h!]
\centering
\includegraphics[scale=0.5]
{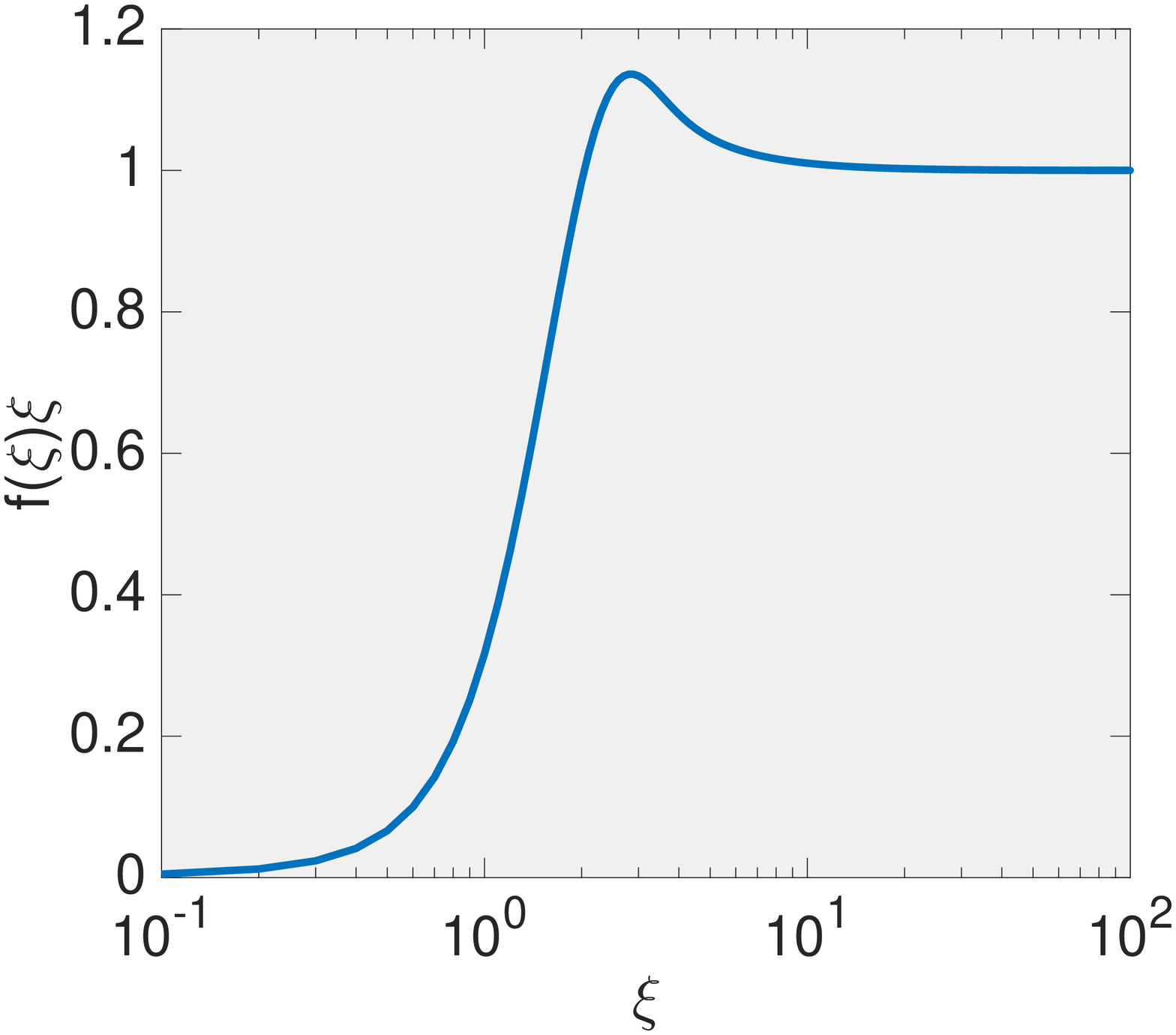}
\caption{\small The numerical evaluation of the second term in Eq. (\ref{eq:AbCh5SI}) for different eigenstate energies.}
\label{fig:EnDep}
\end{figure}

\subsection{Investigation of the localization threshold for Floquet states using a Green function method}

To obtain an accurate estimate of the localization threshold it is convenient to use the Green function method \cite{AbouChacra73}. The Green functions $G^{ab}_{mn}(\epsilon)$ can be introduced using the generalized equation (common dependence of quasi-energy is skipped in all Green function notations)
\begin{eqnarray}
(\epsilon-m\hbar\omega-\Phi_{a})G^{ab}_{mp} = \Gamma\sum_{k,n}J_{m-n}(\delta\sigma_{a k}/\omega)G^{kb}_{np}+\delta_{ab}\delta_{mp}, 
\label{eq:GFSM}
\end{eqnarray}
where indices $a$, $b$, $k$ enumerates the states with fixed spin projections to the $z$ axis, each state $k$ ($k=1,..K$) differ from the state $a$ by the flip of a single spin $k$ with the projection $\sigma_{a k}$ in the state $a$ and $m$, $n$, $p$ are Floquet state indices. Below the case $\delta>\omega$ is always considered since in this case the periodic drive significantly affects the localization. 

\subsubsection{Derivation of the self-consistent equation in the diagonal approximation}

Below the self-consistent equation for the Green functions is derived using methods of Refs. \cite{Anderson58,AbouChacra73}. The equation for the Green function diagonal with respect to the site $a$, $G^{a}_{mp}= G^{aa}_{mp}$, can be written as 
\begin{eqnarray}
(\epsilon+m\hbar\omega-\Phi_{a})G^{a}_{mp}
= \delta_{mp}+\Gamma\sum_{k, n}J_{m-n}(\delta\sigma_{a k}/\omega)G^{ka}_{np}.  
\label{eq:DiagGFSM}
\end{eqnarray}
Equation for Green functions in the right hand side of Eq. (\ref{eq:DiagGFSM}) can be expressed similarly as 
\begin{eqnarray}
(\epsilon+m\hbar\omega-\Phi_{k})G^{ka}_{mp}
= \Gamma\sum_{n}J_{m-n}(-\delta\sigma_{a k}/\omega)G^{a}_{np}
+ \Gamma\sum_{k',n}J_{m-n}(\delta\sigma_{a k'}/\omega)G^{kk',a}_{np},  
\label{eq:FloqGF1SM}
\end{eqnarray}
where states $kk'$ are obtained from the state $a$ using two spin $k$ and $k'$ flips.  

Considering the first sum in the right hand side of Eq. (\ref{eq:FloqGF1SM}) as inhomogeneity one can formally solve this equation for $G^{ka}_{mp}$ using the Green functions $G^{kk}_{mn}$ (in the Bethe lattice problem this Green function is taken for connectivity $K-1$  \cite{AbouChacra73} which should not be important for $K\gg 1$) as  
\begin{eqnarray}
G^{ka}_{mp}=\Gamma \sum_{n, n'}G^{k}_{mn}J_{n-n'}(-\delta\sigma_{a k}/\omega)G^{a}_{n'p}.   
\label{eq:FloqGF2SM}
\end{eqnarray}
This self-consistent approximations neglects the interference of different paths (spin flips taken in different orders). The interference can be neglected in the case of a large coordination number $K$ \cite{Laumann14,AltshullerGefen97,Anderson58} (large number of spins $N$) where the forward approximation is applicable. 

Substituting Eq. (\ref{eq:FloqGF2SM}) into the right hand side of Eq. (\ref{eq:FloqGF1SM}) one obtains the self-consistent equation for the single site Green function 
\begin{eqnarray}
(\epsilon-m\hbar\omega-\Phi_{a})G^{a}_{mp} 
= \delta_{mp}+\Gamma^2\sum_{k,n, n',p'}J_{m-n}(\delta\sigma_{a k}/\omega)G^{k}_{nn'}
J_{n'-p'}(-\delta\sigma_{a k}/\omega)G^{a}_{p'p}. 
\label{eq:FloqGFS3}
\end{eqnarray}

Below this self-consistent equation is analyzed  in the case of a large driving field amplitude and yet sufficiently small field quantization energy satisfying the conditions (remember that $\delta > \omega$)
\begin{eqnarray}
\hbar\delta > \frac{W}{K},  ~~ \hbar\omega < W.   
\label{eq:NonAdS2}
\end{eqnarray}
The case of large quantization energies, $\hbar\delta>\hbar\omega > W$, is considered separately in Sec. \ref{sec:LargeFreq} and the case of small fields is considered in Sec. \ref{sec:SmDelta}. It is shown below that if the inequalities in Eq. (\ref{eq:NonAdS2}) are satisfied then 
one can restrict the consideration in Eq. (\ref{eq:FloqGFS3}) to the only Green functions diagonal with respect to Floquet indices in the vicinity of the localization threshold, $\Gamma_{c}$. The consideration is made in the vicinity of the localization threshold estimated in the main text and below as 
\begin{eqnarray}
\Gamma_{c} \sim \sqrt{\frac{\omega}{\delta}}\frac{W}{K}, ~ \frac{W}{K}<\hbar\delta < W
\label{eq:LocThr1}
\end{eqnarray}
or as 
\begin{eqnarray}
\Gamma_{c} \sim \frac{\sqrt{\omega\delta}}{K}, ~ \hbar\omega<W<\hbar\delta.
\label{eq:LocThr2}
\end{eqnarray}
The  diagonal approximation is a natural extension of a forward approximation  which is expected to be valid in  the   non-adiabatic regime in the Bethe lattice with the large connectivity, where it represents the first non-vanishing contribution to the imaginary part of the self-energy (cf. Ref. \cite{ab13LZTh}). Indeed, in both cases of interest Eqs. (\ref{eq:LocThr1}), (\ref{eq:LocThr2}), level crossing induced by the periodic field occurs in a non-adiabatic Landau-Zener regimes.

One can represent the formal solution of Eq. (\ref{eq:FloqGFS3}) for the Green function  as  
\begin{eqnarray}
\widehat{G}^{a} =\left[(\epsilon-\Phi_{a})\widehat{I}-\hbar\omega \widehat{n} - \widehat{V} \right]^{-1}
\label{eq:FloqGFS4}
\end{eqnarray}
with the operator $\widehat{V}$ defined by its matrix elements in a Floquet indices representation as 
\begin{eqnarray}
V_{mp}=\sum_{k,n,n'}J_{m-n}(\delta\sigma_{a k}/\omega)G^{k}_{nn'}J_{n'-p}(-\delta\sigma_{a k}/\omega).
\label{eq:OperVSM}
\end{eqnarray}
where $\widehat{I}$ and $\widehat{n}$ are diagonal unity matrix and number operator, respectively, in the Floquet state representation. The off-diagonal operator $\widehat{V}$ possesses the symmetry $V_{mm'}=V_{m'm}$ that is the consequence of the properties of Bessel functions with integer indices $J_{n}(x)=J_{-n}(-x)$ \cite{GradshteynRyzhik07}. Therefore one can use the standard perturbation theory to evaluate the Green functions.  

The diagonal Green function can be approximately evaluated using the perturbation theory with respect to the off-diagonal perturbation $\widehat{V}$.  Off-diagonal Green functions in the first non-vanishing order in $\Gamma$ can be expressed using Eq. (\ref{eq:FloqGFS4}) as 
\begin{eqnarray}
\widehat{G}^{a}_{mm'} =G^{a}_{m}V_{mm'}G^{a}_{m'}, ~ m\neq m',
\label{eq:FloqGFoffdS}
\end{eqnarray}
where the notation is introduced $G^{a}_{m}=G^{a}_{mm}$ for fully diagonal Green functions. 
Since off-diagonal Green functions  contain the small perturbation $\widehat{V}$ one can leave only diagonal Green functions in the definition of matrix elements $V_{mm'}$ which yields 
\begin{eqnarray}
\widehat{G}^{a}_{mm'} \approx G^{a}_{m}G^{a}_{m'}\sum_{k,n}J_{m-n}(\delta\sigma_{a k}/\omega)J_{n-m'}(-\delta\sigma_{a k}/\omega)G^{k}_{n}.
\label{eq:FloqGFoffdS1}
\end{eqnarray}

One can represent the diagonal Green function in the standard form introducing the self-energy $\Sigma^{a}_{m}$ as
\begin{eqnarray}
G^{a}_{m}=\frac{1}{\epsilon-m\hbar\omega-\Phi_{a}-\Sigma^{a}_{m}}.
\label{eq:FloqGFDiagS}
\end{eqnarray}
The self-energy $\Sigma^{a}_{m}$ can be expanded up to the second order of perturbation theory in the form  
\begin{eqnarray}
\Sigma^{a}_{m}=V_{mm}+\sum_{n\neq m}\frac{V_{mn}^2}{\hbar\omega(m-n)}.   
\label{eq:FloqSES}
\end{eqnarray}
Since off-diagonal Green functions contain small matrix elements of the perturbation $\widehat{V}$ (cf. Eq. (\ref{eq:FloqGFoffdS})) the first and second order (in $V$) contributions to the self-energy can be represented as 
\begin{eqnarray}
\Sigma^{a}_{m}=\Sigma^{a}_{m,1}+\Sigma^{a}_{m,2},
\nonumber\\
\Sigma^{a}_{m,1}=\Gamma^2\sum_{k,n}J_{m-n}(\delta/\omega)^2G^{k}_{n},
\nonumber\\
\Sigma^{a}_{m,2}=\Gamma^2\sum_{k,n,n' (n\neq n')}J_{m-n}(\delta\sigma_{ k}/\omega)J_{n'-m}(-\delta\sigma_{a k}/\omega)G^{k}_{nn'}
\nonumber\\
+ 
\Gamma^4\sum_{k, n (n\neq m)}\frac{\left(\sum_{p}G^{k}_{p}J_{m-p}(\delta\sigma_{a k}/\omega)J_{p-n}(-\delta\sigma_{a k}/\omega)\right)^2}{\hbar\omega(m-n)}.  
\label{eq:FloqSES1}
\end{eqnarray}
Below it is demonstrated that the second correction, $\Sigma^{a}_{m,2}$, can be neglected compared to the first order term if Eq. (\ref{eq:NonAdS2}) is satisfied. In this case the equation for the consideration is restricted to diagonal Green functions only. 

To estimate the second order correction consider the first contribution to $\Sigma^{a}_{m,2}$, Eq. (\ref{eq:FloqGFoffdS1}) (the second term can be treated similarly and it is comparable to the first term). This contribution should be considered separately for intermediate, Eq. (\ref{eq:LocThr1}), and large Eq. (\ref{eq:LocThr2}) driving fields. 

In the case of an intermediate driving field the second order correction includes the sum of contributions of $K$ possible flips of different spins $k$ and there are $n_{F}=\delta/\omega$ significant Floquet states for each spin so the total number of terms is given by $Kn_{F}$ 
(remember that $J_{n}(\delta/\omega) \sim \sqrt{\omega/\delta}$ for $n< n_{F}=\delta/\omega$ and this function can be approximately neglected for larger $n$).   The typical value for the minimum denominator of the Green function $G^{k}_{n}$ in the sum in Eq. (\ref{eq:FloqGFoffdS1}) can be estimated as $W/(Kn_{F})$ and the term with the minimum denominator gives a reasonable estimate for the typical value of the whole sum (see e. g. the analysis of Ref.  \cite{Raikh00Levy}) in the form 
\begin{eqnarray}
\widehat{G}^{a}_{m|m'} \sim \eta_{mm'}G^{a}_{m}G^{a}_{m'}\frac{K\Gamma^2}{W},
\label{eq:FloqGFoffdS2}
\end{eqnarray}
where $\eta_{mm'}$ is a sign variable function of order of unity. 
Substituting this result into the first term  of the second order correction to the self-energy yields
\begin{eqnarray}
\frac{K\Gamma^2}{W}\Gamma^2\sum_{k,n,n' (n\neq n')}\eta_{nn'}J_{m-n}(\delta\sigma_{a k}/\omega)J_{n'-m}(-\delta\sigma_{a k}/\omega)G^{k}_{n}G^{k}_{n'}.  
\label{eq:FloqSES3}
\end{eqnarray}
 The sum is determined by the minimum denominator ($|\epsilon-\Phi_{k}-s\hbar\omega| \leq W/(Kn_{F})$) in one out of two Green functions in Eq. (\ref{eq:FloqSES3}) (either $s=n$ or $s=n'$).  Moreover the most important contribution to the localization threshold comes from the domain  $|\epsilon-\Phi_{k}-s\hbar\omega| \leq \omega$. In that case the absolute value of the second Green function can be estimated as $1/(\hbar\omega|n-n'|)$. Since the series in Eq. (\ref{eq:FloqSES3}) is sign variable the main contribution comes from the terms with $n'-n\sim 1$. Then the expression in Eq. (\ref{eq:FloqSES3}) can be estimated as 
 \begin{eqnarray}
\frac{K\Gamma^2}{W\hbar\omega}\times \Gamma^2\sum_{k,n}\eta_{nn'}J_{m-n}(\delta\sigma_{a k}/\omega)J_{n-m+1}(-\delta\sigma_{a k}/\omega)G^{k}_{n} \sim \frac{K\Gamma^2}{W\hbar\omega}\times \Sigma^{ak}_{m,1},  
\label{eq:FloqSES4}
\end{eqnarray}
 since the second factor in the product possesses a similar structure to the first order correction to the self-energy in Eq. (\ref{eq:FloqSES1}). 
 
The first factor, $K\Gamma^2/(W\hbar\omega)$, in Eq. (\ref{eq:FloqSES4}) estimates the relative value of the second order correction to the self-energy and if it is much less then unity then this correction can be neglected. Using Eq. (\ref{eq:LocThr1})  one gets $K\Gamma^2/(W\hbar\omega) \sim (\Gamma/\Gamma_{c})^2 W/(K\hbar\delta) \ll 1$, which proves the weakness of the second order correction for intermediate fields. 

The similar consideration can be applied to large fields, $\hbar\delta > W$. In this case the number of terms contributing to the second order correction, Eq. (\ref{eq:FloqSES1}), is given by $KW/(\hbar\omega)$ since the probability to find the resonant state with random energy exceeding $W$ is exponentially small. Consequently, the first order approximation in Eq. (\ref{eq:FloqSES1}) is valid if the inequality $K\Gamma^2/(\hbar^2\omega\delta) \ll 1$ is satisfied near the localization threshold. According to Eq. (\ref{eq:LocThr2})  this inequality is always satisfied near the localization threshold at $K \gg 1$. 

Notice that in both cases one can express the ratio of the second order correction to the first order term in the form of the product $(\Gamma/\Gamma_{c})\times (\Gamma/(\hbar\sqrt{\omega\delta}))$. The first factor in the product is of of order of $1$ near the transition point while the second factor is much smaller then $1$ for non-adiabatic level crossings in a full accord with the initial expectation that the perturbation theory expansion is applicable in the non-adiabatic regime \cite{ab13LZTh}. 

If  the second order correction to the self-energy in Eq. (\ref{eq:FloqSES1}) can be neglected compared to the first order contribution $\Sigma^{a}_{m,1}$ one can use  the diagonal approximation suggesting that 
\begin{eqnarray}
G^{a}_{m}=\frac{1}{\epsilon-m\hbar\omega-\Phi_{a}-\Gamma^2\sum_{k,n}J_{m-n}(\delta/\omega)^2 G^{k}_{n}}.  
\label{eq:FloqGFDiag}
\end{eqnarray}
Consequently the imaginary part of the self-energy obeys   the self-consistent equation 
\begin{eqnarray}
{\rm Im} \Sigma^{a}_{m}=\Gamma^2\sum_{k, n}|G^{k}_{n}|^2 J_{m-n}(\delta/\omega)^2 {\rm Im}\Sigma^{k}_{n}  
\label{eq:FloqSEDiag}
\end{eqnarray}
and the self-energy $\Sigma^{a}_{m}$ is independent of random energy $\Phi_{a}$ of the state $a$ in the Bethe lattice approximation.

\subsubsection{Analysis of the localization threshold}

The delocalization transition takes place at the critical transverse field $\Gamma_{c}$ where Eq. (\ref{eq:FloqSEDiag}) acquires a non-zero solution for imaginary parts of self-energies. To estimate this field one can employ the procedure of Ref. \cite{AbouChacra73} to Eq. (\ref{eq:FloqSEDiag}) as 
\begin{eqnarray}
\left<\exp(-t{\rm Im} \Sigma^{a}_{m})\right>
=\left<\exp\left(-t\Gamma^2\sum_{n}|G^{k}_{n}|^2 J_{m-n}(\delta/\omega)^2 {\rm Im}\Sigma^{k}_{n}\right)\right>^{K},
\label{eq:FloqSEAbCh1}
\end{eqnarray}
where averaging is performed over all random energies $\Phi$ and the parameter $t$ is greater than $0$ and considered in the limit $t \rightarrow 0$. The left hand side in Eq. (\ref{eq:FloqSEAbCh1}) can be taken in the form \cite{AbouChacra73,ab16preprintSG} 
$\left<\exp(-t{\rm Im} \Sigma^{a}_{m})\right>\approx 1-\sqrt{t}F_{m}(\epsilon)$.
To evaluate the right hand side in Eq. (\ref{eq:FloqSEAbCh1})  one can approximate the Green functions there by their zeroth order expressions $G^{k}_{n} \approx 1/(\epsilon-n\hbar\omega-\Phi_{k})$. This is well justified in the case of a large coordination number $K \gg 1$ if the denominator is not extremely small so the associated correction to the real part of the self-energy is smaller than the typical random energy, $\Gamma^2 J_{m-n}(\delta/\omega)^2/|\epsilon-n\hbar\omega-\Phi_{k}| \ll W$ \cite{AbouChacra73}. 

In the case of interest, Eq. (\ref{eq:Limits00}), the exponent in the right hand side of Eq. (\ref{eq:FloqSEAbCh1}) strongly depends on a random energy $\Phi_{k}$. 
Since all imaginary parts of self-energies for $n<n_{F}$ are of the same order of magnitude (see Eq. (\ref{eq:FloqSEDiag})) the main contribution to the exponent in the right hand side of Eq. (\ref{eq:FloqSEAbCh1}) at a certain random energy $\Phi_{k}$ is given by the closest $n^{th}$ resonance  $|\epsilon-\Phi_{k}-n\hbar\omega)| < \hbar\omega/2$. 
Then one can  replace the sum in exponent in Eq. (\ref{eq:FloqSEAbCh1}) with the contribution of those resonances expressing the integral over a random energy in the right hand side of Eq. (\ref{eq:FloqSEAbCh1}) as the sum of resonant contributions from the domains $|\epsilon-\Phi_{k}-n\hbar\omega)| < \hbar\omega/2$. Each contribution can be evaluated as \cite{AbouChacra73}
\begin{eqnarray}
-\sqrt{t}\Gamma F_{n}(\epsilon)|J_{m-n}(\delta/\omega)|\int_{\epsilon-(n+1/2)\hbar\omega}^{\epsilon-(n-1/2)\hbar\omega}\frac{P(\Phi)d\Phi}{|\epsilon-n\hbar\omega-\Phi|}\theta(|\epsilon-n\hbar\omega-\Phi|-\Gamma^2J_{m-n}(\delta/\omega)^2/W).
\label{eq:FloqSEAbCh4a}
\end{eqnarray}
The Heaviside theta function is introduced to avoid corrections to a real part of self-energy exceeding the typical energy $W$ \cite{AbouChacra73}. Then the logarithmic integral can be approximately evaluated as   $\ln(W\hbar\omega/(\Gamma^2J_{m-n}(\delta/\omega)^2)$ and near the localization thresholds, Eqs. (\ref{eq:LocThr1}), (\ref{eq:LocThr2}), one can replace the argument of logarithm by the common expression $\xi=\frac{K^2\hbar\delta^2 W}{\omega(W^2+(\hbar\delta)^2)}$ covering both regimes with the logarithmic accuracy. 
This results in the self-consistent definition of the localization threshold by the equation 
\begin{eqnarray}
F_{m}(\epsilon)=2\ln(\xi)K\Gamma \sum_{n} |J_{m-n}(\delta/\omega)|P(\epsilon-\hbar\omega n)
F_{n}(\epsilon).
\label{eq:FloqSEAbCh4}
\end{eqnarray}
The delocalization takes place when Eq. (\ref{eq:FloqSEAbCh4}) acquires a first non-zero solution. 

The numerical solution of Eq. (\ref{eq:FloqSEAbCh4}) has been used in the main text to calculate dependencies of the localization threshold shown in Figs. 2 and 3 in the case $\hbar\delta>W/K$, $\hbar\omega<W$ and to sketch Figs. \ref{fig:BesselAv}, \ref{fig:Thr}, \ref{fig:PhDiagr}, \ref{fig:GamcOm}  below. 

In either case of $\hbar\delta \ll W$ or $\hbar\delta \gg W$ one can assume $F_{m}(\epsilon)$ to be approximately independent of a Floquet index $m$ at small $m$ which defines the localization threshold as
\begin{eqnarray}
\Gamma_{c}=\frac{1}{2K\ln(\xi)\sum_{m}|J_{m}(\delta/\omega)|P(\epsilon-m\hbar\omega)}.
\label{eq:AnsAnalit}
\end{eqnarray}
The sum in Eq. (\ref{eq:AnsAnalit}) can be evaluated analytically in different asymptotic limits as derived below. The asymptotic  results are consistent with the numerical solution of Eq. (\ref{eq:FloqSEAbCh4}) shown in Figs. 2 and 3 in the main text and in Figs. \ref{fig:Thr}, \ref{fig:BesselAv} here in Supplementary Materials.

\begin{figure}[h!]
\centering
\includegraphics[scale=0.5]
{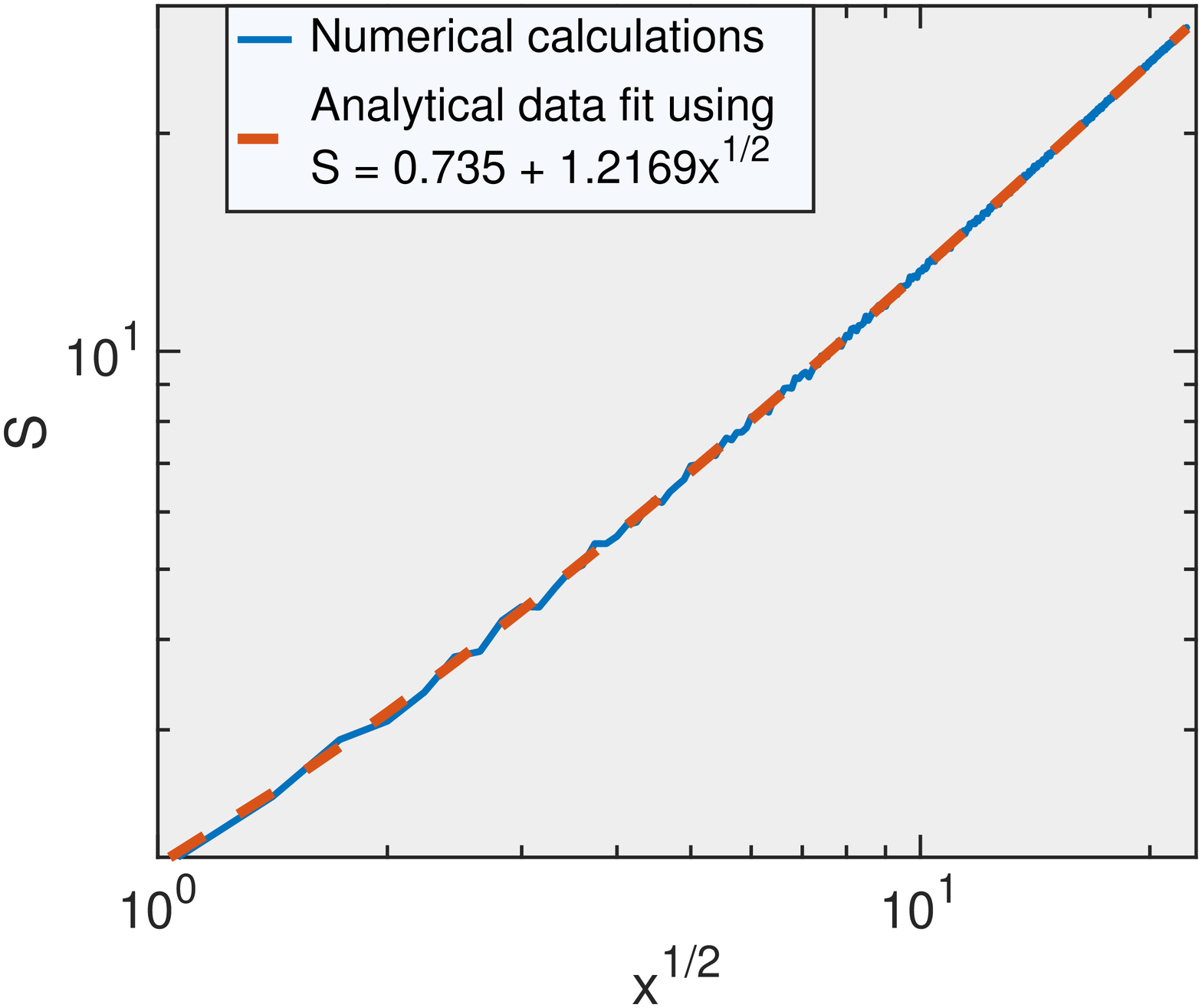}
\caption{\small Analytical asymptotic fit of the infinite sum of absolute values of Bessel functions $S(x)=\sum_{m}|J_{m}(x)|$. }
\label{fig:BesselSum}
\end{figure} 
 
\subsubsection{Asymptotic analytical behaviors of localization threshold}

Consider the solutions of Eq. (\ref{eq:AnsAnalit}) in the limits $\hbar\delta < W$ and $\hbar\delta > W$ separately for $\epsilon=0$. In the first case the function $P$ changes very smoothly with $n$ on the scale of the significant change of the Bessel function $n\hbar\omega \sim \hbar\delta \ll W$ so it can be approximately replaced with the constant $P(0)$. Then the localization threshold can be estimated as 
\begin{eqnarray}
\Gamma_{c}=\frac{1}{2P(0)\ln(\xi)\sum_{m}|J_{m}(\delta/\omega)|}, ~ \xi=\frac{\hbar\delta^2}{\omega(W+\hbar\delta)}.
\label{eq:LocThr1SM}
\end{eqnarray} 
According to the Bessel function asymptotic behavior \cite{GradshteynRyzhik07}
at large argument one can expect that the infinite sum has $\delta/\omega$ contributions of order of $\sqrt{\omega/\delta}$ so it can be estimated as $\sqrt{\delta/\omega}$. The numerical analysis (see Fig. \ref{fig:BesselSum}) gives the more accurate estimate 
\begin{eqnarray}
\sum_{m} |J_{m}(x)| \approx 0.735+1.2169\sqrt{\frac{\delta}{\omega}}, 
\label{eq:BesselSumEstSM}
\end{eqnarray} 
used within the main text.

\begin{figure}[h!]
\centering
\includegraphics[scale=0.5]
{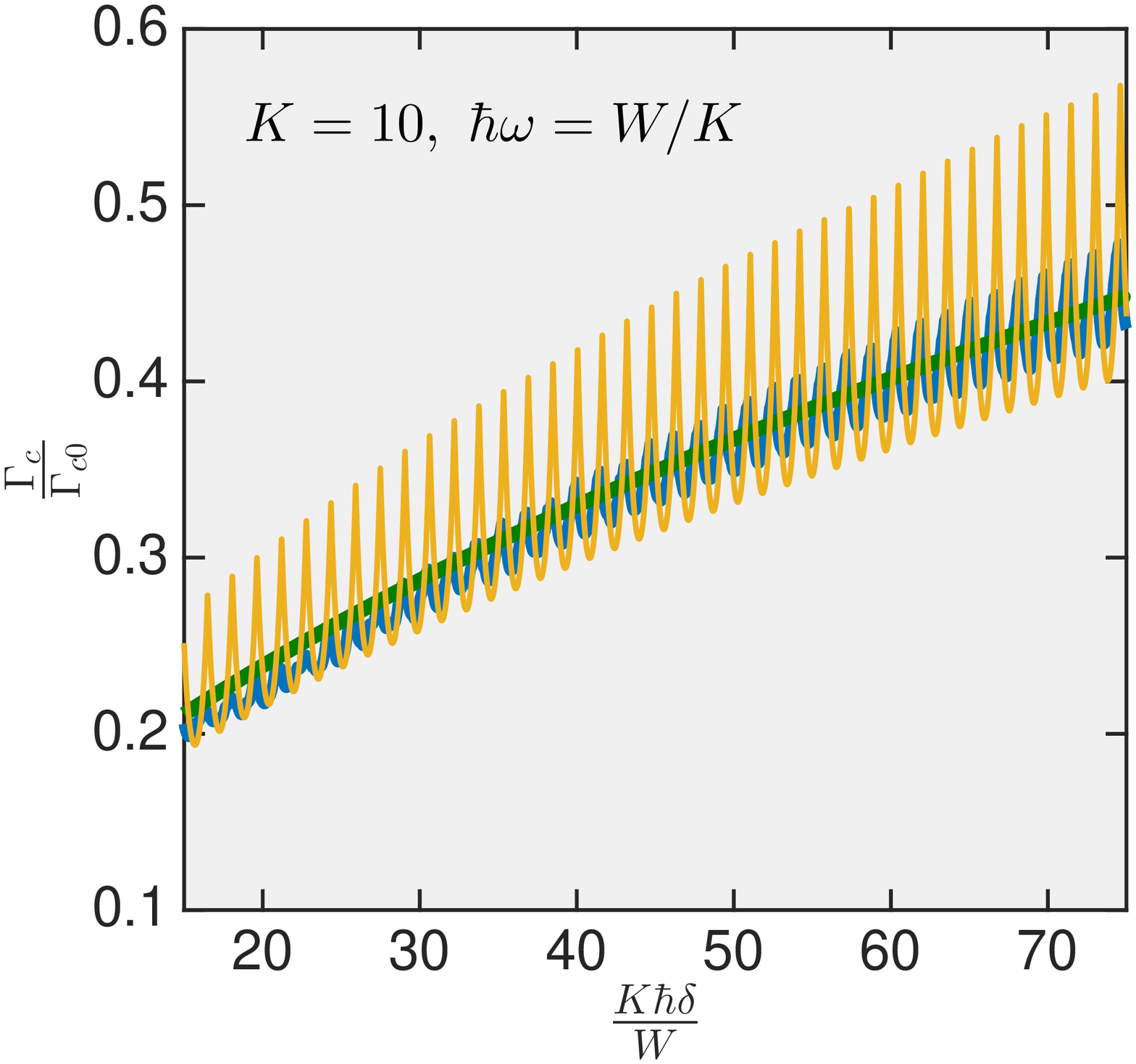}
\caption{\small The drive induced modification of the localization threshold vs. the drive amplitude in the case $\hbar\omega < W < \hbar\delta$.  The numerical solution of Eq. (\ref{eq:FloqSEAbCh4}) for $\Gamma_{c}/\Gamma_{c0}$  shown by the blue line (oscillating with the smallest amplitude) is compared to the average solution, Eq. (\ref{eq:BesselAsymptAvSM}), shown by the green line (nearly straight) and the solution obtained using the Bessel function asymptotic behavior, Eq. (\ref{eq:BesselAsymptAvSMA}), shown by the yellow line (oscillating with the largest amplitude). }
\label{fig:BesselAv}
\end{figure} 
 
In the opposite limit of $\hbar\delta > W$ (remember that $\epsilon=0$) the sum in the denominator of Eq. (\ref{eq:AnsAnalit}) is determined by Bessel functions of  order $m\ll \delta/\omega$. Then one can approximately replace Bessel functions, $J_{m}(x)$,  in Eq. (\ref{eq:AnsAnalit})  with their average value with respect to their order $<|J_{m}(\delta/\omega)|>_{m}$. Consequently, the localization threshold can be defined as 
\begin{eqnarray}
\Gamma_{c}=\frac{1}{2K\ln(\xi)<|J_{m}(\delta/\omega)|>_{m}\sum_{m}P(m\hbar\omega)} \approx \frac{\hbar\omega}{2K\ln(\xi)<|J_{m}(\delta/\omega)|>_{m}}, ~ N\gg 1.
\label{eq:LocThr2SM}
\end{eqnarray} 

The average Bessel function can be estimated using its asymptotic behavior at large argument \cite{GradshteynRyzhik07}
\begin{eqnarray}
J_{m}(x) \approx \sqrt{\frac{2}{\pi x}}\cos(x-m\pi/2-\pi/4), ~ x\gg m. 
\label{eq:BesselAsymptSM}
\end{eqnarray} 
To characterize the general trend in the field and frequency dependencies of the localization threshold one can perform continuous  averaging of the asymptotic expression Eq. (\ref{eq:BesselAsymptSM}) with respect to its index $m$ as 
\begin{eqnarray}
<|J_{m}(x)|>_{m}=\sqrt{\frac{2}{\pi x}}\frac{1}{M}\int_{-M/2}^{M/2}|\cos(x-m\pi/2-\pi/4)|dm=\frac{(2/\pi)^{3/2}}{\sqrt{x}}, ~ 1 \ll M \ll x. 
\label{eq:BesselAsymptAvSM}
\end{eqnarray} 
More sophisticated approximation can be constructed using explicitly the analytical dependence in Eq. (\ref{eq:BesselAsymptSM}). This yields 
\begin{eqnarray}
<|J_{m}(x)|>_{m} \approx \frac{\sqrt{1+|\cos(2x)|}}{\sqrt{2\pi x}}.  
\label{eq:BesselAsymptAvSMA}
\end{eqnarray}
In Fig. \ref{fig:BesselAv} the numerical solution of Eq. (\ref{eq:FloqSEAbCh4}) is compared to the average and accurate asymptotic solutions. The match between numerical and asymptotic oscillations clearly exists, but it is far from  perfect possibly because of an inaccuracy of the asymptotic expression in Eq. (\ref{eq:BesselAsymptAvSM}). Therefore the average solution is used in the main text to represent the localization threshold behavior.
 
 In Fig. \ref{fig:Thr} the numerical solution of the self-consistent equation, Eq. (\ref{eq:FloqSEAbCh4}), is compared to the asymptotic analytical approximations for two different relationships between the characteristic spin flip energy and the field quantization energy. It is clear that the asymptotic expressions describe the solution quite accurately except for the crossover regime $\hbar\delta \approx W$. 
 
\begin{figure}[h!]
\centering
\includegraphics[scale=0.5]{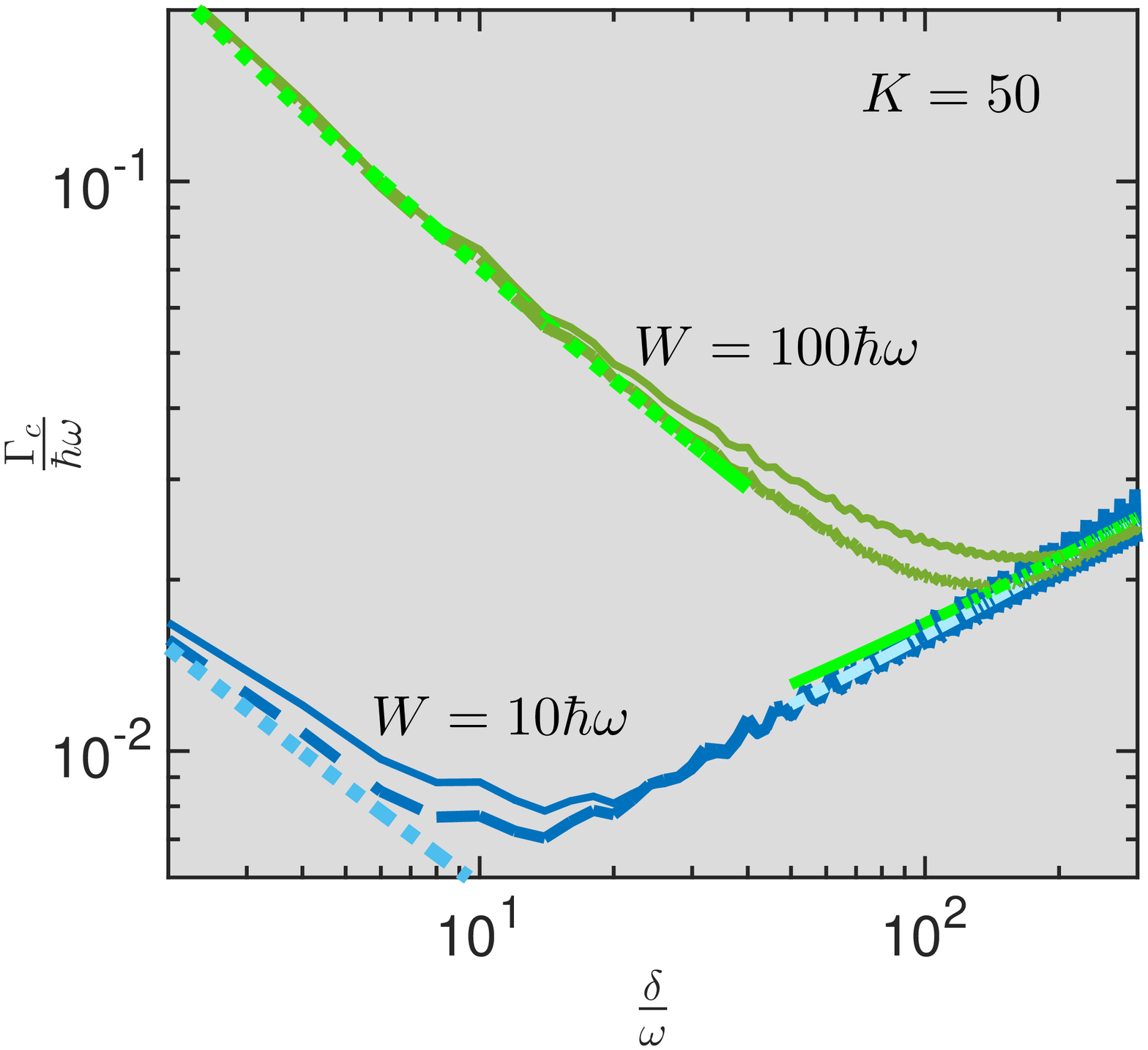}
\caption{\small Localization threshold dependence on the driving field. Straight lines show numerical solutions of Eq. (\ref{eq:FloqSEAbCh4}) for  $W=10\hbar\omega$ and $100\hbar\omega$, dashed lines show approximate analytical solutions, Eq. (\ref{eq:AnsAnalit}), and dotted lines show asymptotic behaviors (see Eq. (5) in the main text).}
\label{fig:Thr}
\end{figure}

\subsection{Localization threshold at high frequency $\hbar\omega > W$ near Bessel function zeros}
\label{sec:LargeFreq}

In the case of a large quantization energy  compared to a random energy, $\hbar\omega \gg W$, the coupling of states with different Floquet indices $n$ ($\Gamma \sqrt{\omega/\delta}$) is much smaller than their energy splitting, $\hbar\omega$, near the localization threshold in accordance with the localization threshold estimate in this regime. Consequently it can be treated as a small perturbation compared to the coupling $\Gamma J_{0}(\delta/\omega)$ conserving the Floquet index provided that the ratio $\delta/\omega$ is not close to one of the zeros of the Bessel function. Then the localization threshold can be estimated as $\Gamma_{c} \approx \Gamma_{c0}/|J_{0}(\delta/\omega)| \sim \sqrt{\delta/\omega}\Gamma_{c0}$ where $\Gamma_{c0}$ is the threshold in the absence of the drive. 

In the case of strongly suppressed coupling, $J_{0}(\delta/\omega) \rightarrow 0$, virtual transitions through states having a non-zero Floquet index $n$  should be included perturbatively to generate a non-zero coupling. In the first non-vanishing approach the perturbation theory  generates the coupling to the second coordination sphere (states obtained from the state $a$ flipping two spins $k$ and $k'$), which can be estimated as (see e. g. Ref. \cite{ab15MBLXY})
\begin{eqnarray}
\Gamma_{2}^{kk'} = \Gamma^2\sum_{n=-\infty}^{\infty}\left[\frac{J_{n}(\sigma_{ka}\delta/\omega)J_{-n}(\sigma_{k'a}\delta/\omega)}{n\hbar\omega +\Phi_{k}-\Phi_{a}}+\frac{J_{n}(\sigma_{k'a}\delta/\omega)J_{-n}(\sigma_{ka}\delta/\omega)}{n\hbar\omega +\Phi_{k'}-\Phi_{a}}\right].  
\label{eq:BessZeroS}
\end{eqnarray} 
If random energy terms ($\Phi_{k}-\Phi_{a}$) are ignored in the denominator compared to the quantization energies then the expression is equal zero because the replacement $n\rightarrow -n$ changes the sign of the term to the opposite one. Therefore the expansion should be made with respect to the small energy difference which leads to the estimate 
\begin{eqnarray}
\Gamma_{2}^{kk'} = \frac{2\Gamma^2(\Phi_{k}+\Phi_{k'}-2\Phi_{a})}{(\hbar\omega)^2}\sum_{n=1}^{\infty}\frac{(-1)^{n}(\sigma_{ka}\sigma_{k'a})^{n}|J_{n}(\delta/\omega)|^{2}}{n^2}.  
\label{eq:BessZero1S}
\end{eqnarray} 

Using the Bessel function asymptotic expression Eq. (\ref{eq:BesselAsymptSM}) and the Euler formula $\sum_{n=1}^{\infty}\frac{1}{n^2}=\frac{\pi^2}{6}$ one can evaluate the coupling energies in Eq.  (\ref{eq:BessZero1S}) at Bessel function zeros as 
\begin{eqnarray}
\Gamma_{2}^{kk'} = -\frac{\pi}{2}\frac{\sigma_{ka}\sigma_{k'a}\Gamma^2(\Phi_{k}+\Phi_{k'}-2\Phi_{a})}{\hbar\omega\delta}.  
\label{eq:BessZero2S}
\end{eqnarray} 
This result is consistent with the numerical evaluation of the sums in Eq. (\ref{eq:BessZero1S}) evaluated in Bessel function zeros taken from Ref. \cite{BesselZeros}.


Consequently in the case $J_{0}(\delta/\omega)=0$ the localization threshold can be  determined by  the alternative problem for the Bethe lattice with the coupling $\Gamma_{2}$, coordination number $K^{2}$ and characteristic random energy distribution $P \sim 1/W$. The localization threshold  for the quasi-energy equal to zero can be then estimated using the equation \cite{AbouChacra73} 
\begin{eqnarray}
1 = 4K^2 P(0)<|\Gamma_{2}^{kk'}|>\ln(K^2)= 4\sqrt{6}K^2 \frac{\Gamma^2}{\hbar^2\omega\delta}  \ln(K). 
\label{eq:BessZero3S}
\end{eqnarray} 
The localization threshold in the case of the ratio $\delta/\omega$ corresponding to the  Bessel function zero can be expressed as 
\begin{eqnarray}
\Gamma_{c,pert}=\frac{\hbar\sqrt{\omega\delta}}{2K6^{1/4}\sqrt{\ln(K)}}.
\label{eq:BessZero4S}
\end{eqnarray} 
This expression gives the upper constraint for the localization threshold near the Bessel function zeros. It exceeds the ``typical" localization threshold $\sqrt{\delta/\omega}W/N$ by the factor $\hbar\omega/W$, which can be potentially made much greater than unity. The common expression for the localization threshold in the case of a large frequency $\omega > W/\hbar$ can be summarized in the form 
\begin{eqnarray}
\Gamma_{c} \approx \frac{\Gamma_{c0}}{{\rm max} \left(|J_{0}(\delta/\omega)|, ~ \frac{6^{1/4}\sqrt{\pi}W}{\hbar\sqrt{2\omega\delta\ln(K)}}\right)},   
\label{eq:HiFreqFinSM}
\end{eqnarray} 
used in the main text. 
The quasi-energy dependence of the localization threshold can be described following the derivation in Sec. \ref{sec:NoDrive}, Eq. (\ref{eq:AbCh6SI}).  The localization threshold at a finite quasi-energy $\epsilon$ can be determined replacing $\Gamma_{c0}$ with $\Gamma_{c0}(\epsilon)$ in Eq. (\ref{eq:HiFreqFinSM}).

\subsection{Case of a small driving field $\hbar\delta < W/K$.} 
\label{sec:SmDelta}
 
Here some detail of the localization threshold estimate in the case of a small driving field, $\hbar\delta \ll W/K$, and low frequency, $\omega \ll \delta$, are given, which are complementary to the main text derivation. The localization threshold is estimated  setting the number of resonant interactions  of the given product state (the state with given spin projections to the $z$ axis)  with other states to unity. Assume that the system is localized in the absence of a periodic drive so the delocalization can be only due to resonant interactions induced by this drive. 

Consider the probability of a resonant interaction for two coupled states in the presence of a periodic sinusoidal field.  The effective Hamiltonian for such coupling can be written as 
\begin{eqnarray}
\widehat{H}_{p}=-\Delta \sigma^{z} -V\sigma^{x}-\hbar\delta_{*}\cos(\omega t)\sigma^{z},
\label{eq:TwoLevHSISM}
\end{eqnarray}
where $\delta_{*}$ is the effective field amplitude depending on both states $\sigma^{z}=\pm 1$ and $\Delta$ is a random difference of two states diagonal energies with a typical scale $W$.  As in the main text it is assumed that $\delta_{*}\sim \delta > \omega$. 

Since the system is localized in the absence of a driving field one can ignore its resonances and assume, $\Delta>V$. If one considers the coupling, $V$, as a perturbation then the effective coupling between nearly resonant Floquet states is given by $VJ_{k}(\delta_{*}/\omega)$ where $k \approx \Delta/(\hbar\omega)$ estimates the number of quanta needed to compensate the energy splitting of two states. In the case $\Delta > \hbar\delta_{*}$ (remember, that $\omega \ll \delta$) the Bessel function is exponentially small ($J_{k}(\delta_{*}/\omega) \sim e^{-|\Delta|\ln(|\delta_{*}|/\omega)/(\hbar\omega)}$ \cite{GradshteynRyzhik07}) and one can ignore resonances in this case. Consequently, the only situation where a driving field results in level crossing, $\hbar\delta_{*}\geq \Delta$, is of interest \cite{Abanin16TimeDepRig}. 
In this case the coupling strength can be estimated as \cite{GradshteynRyzhik07} $V_{*}=V\sqrt{\omega/\delta_{*}}$. For a  small coupling, $V_{*} < \hbar\omega$ (non-adiabatic regime), the interaction $V_{*}$ can lead to $\delta_{*}/\omega$ resonances with the total probability $P_{res} \sim P(0)|V|\sqrt{\delta_{d}/\omega}$ where $P(0) \sim 1/W$ is the distribution function of energy differences $\Delta$ in Eq. (\ref{eq:TLResASM}) near zero energy. The assumption of a small coupling will be confirmed by the final estimate of the localization threshold. 

Since the field amplitude is small, $\hbar\delta \ll W/K$, there is a negligible amount of crossings induced by the drive between the given state and the $K$ neighboring states different from the given state by a single spin flip. Level crossings takes place in the $p^{th}$ coordination sphere containing $K^{p}/p!$ states different from the given states by $p$ spin flips, where $Wp!/K^{p} \sim \hbar\delta$. Consequently the number of the relevant coordination sphere can be estimated using the Stirling formula, $\ln(n!)\approx n\ln(n/e)$, as  
\begin{eqnarray}
p \approx \frac{\ln(W/(\hbar\delta))}{\ln(eK/p)}. 
\label{eq:pSM}
\end{eqnarray} 
In the case of interest $p \ll K$ one can approximate this number as $p = \ln(W/(\hbar\delta))/\ln(K)$ and this result, consistent with Ref. \cite{Abanin16TimeDepRig} is used in the main text. Consequently one has to assume $\hbar\delta \gg W/2^{K}$; in the opposite limit the drive amplitude is less than the typical minimum level splitting of the given state with $2^{K}$ states available in a random energy model ($K=N$). In this case the drive is obviously too weak to disturb any localized state. 

The coupling $V$ in Eq. (\ref{eq:TLResASM}) between states separated by $p$ spin flips  is determined by the contribution of $p!$ paths for flips of $p$ spins corresponding to $p!$ permutations in  spin flip orders, which can be expressed as the sum of $p!$ terms  taken within the forward approximation as
\begin{eqnarray}
V_{p}=\sum_{P}\frac{\Gamma^{p}}{\prod_{i=1}^{n-1}(\Phi_{i(p)}-\Phi_{a})}, 
\label{eq:VSpinsSM}
\end{eqnarray} 
where $P$ enumerates permutations and $\Phi_{i(P)}$ is the energy of the state obtained from the initial state by flipping first $i$ spins in the permutation $P$ and $\Phi_{a}$ is the energy of the state under consideration. All denominators in Eq. (\ref{eq:VSpinsSM}) exceeds $\hbar\delta$ since level crossings are expected only for $p$ or more spin flips and therefore intermediate Floquet states are not important. 
The effective field amplitude, $\delta_{*}$, in Eq. (\ref{eq:TLResASM}) for states different by $p$ spin flips can be estimated as $\sqrt{p}\delta$ where $\delta$ describes the longitudinal alternating field introduced in the main text.  

Consequently the total probability of resonance can be estimated summing up all probabilities for $K^{p}/p!$ possible couplings which yields 
\begin{eqnarray}
P_{res,tot} \sim \frac{K^{p}}{p!}\sqrt{\frac{\sqrt{p}\delta}{\omega}}P(0)<|V_{p}|>.  
\label{eq:TLResSM}
\end{eqnarray} 
The localization threshold can be estimated setting this probability to unity. Since near the localization threshold the Floquet state coupling strength $<|V_{p}|>\sqrt{\omega/\delta} \sim \hbar\omega Wp!/(\hbar\delta K^p)$ does not exceed the field quantization energy $\hbar\omega$ different Floquet state resonances  can be. indeed, count as independent. 

Averaging of coupling in Eq. (\ref{eq:VSpinsSM}) over random energies can be performed assuming that the sum there is determined by the largest term \cite{Laumann14,ab16preprintSG} which yields 
\begin{eqnarray}
<|V_{ps}|>\approx\sum_{P}\left<\left|\frac{\Gamma^{p}}{\prod_{i=1}^{n-1}(\Phi_{i(p)}-\Phi_{a})}\right|\right>. 
\label{eq:VSpins1SM}
\end{eqnarray} 
Each term in Eq. (\ref{eq:VSpins1SM}) diverges logarithmically. This logarithmic divergence can be overcome introducing cutoffs for intergrals over energies. The upper  cutoff is given by $|\Phi_{i}-\Phi_{a}| < W$ due to the finite width $W$ of the energy distribution. The lower cutoff is determined in Ref. \cite{AbouChacra73} by the requirement of avoidance of large  contributions to  real parts of self-energies (perturbation theory breakdown) expressed by  the constraint $|\Phi_{i}-\Phi_{a}| > \Gamma^2/W$.  Additional restriction $|\Phi_{i}-\Phi_{a}| > \hbar\delta_{*}$ is required for the validity of the forward approximation   Eq. (\ref{eq:VSpinsSM}); otherwise the intermediate Floquet states should be included into the perturbation theory series. Statistically they cannot be important since the probability of level crossing $|\Phi_{i}-\Phi_{a}| < \hbar\delta_{*}$ with the state located closer than the $p^{th}$ coordination sphere is much smaller than unity. 

Then averaging over intermediate energies in Eq. (\ref{eq:VSpins1SM}) leads to  the equation determining the localization threshold in the form  
\begin{eqnarray}
(2KP(0)\Gamma_{c} L_{1})^{p}\sqrt{\frac{\delta_{*}}{\omega}} =1, ~ L_{1}=\ln(\xi_{1}), ~  \xi_{1}=\frac{K}{\sqrt{\left(\frac{\omega}{\delta}\right)^{\frac{\ln(K)}{2\ln(W/\hbar\delta)}}+\frac{\hbar\delta}{W}}}.
\label{eq:TLResASM}
\end{eqnarray} 
Accordingly the localization threshold can be estimated as 
\begin{eqnarray}
\Gamma_{c}=\frac{1}{2KP(0)L_{1}}\left(\frac{\omega}{\delta}\right)^{\frac{1}{2p}}\approx 
\Gamma_{c0}\left(\frac{\omega}{\delta}\right)^{\frac{1}{2p}}\frac{\ln(K)}{\ln(\xi_{1})}, 
\label{eq:nEstSM}
\end{eqnarray} 
where $\Gamma_{c0}$ stands for the localization threshold in the absence of a periodic drive. This result is used within the main text. 
The factor $p^{1/(4p)}$ has been neglected since it approaches unity for $p>1$.    Although the estimate in Eq. (\ref{eq:nEstSM}) is based on  a simple delocalization criterion that the  number of resonances is of order of one so it skips numerical and/or logarithmic factors \cite{AbouChacra73}, the final answer will contain these factors being raised to the power $1/p$, which makes them negligible in the case $p \gg 1$.

The proposed consideration is applicable only for $p \ll K$ suggesting that the driving field amplitude is much greater than the interstate energy splitting $W/2^{K}$ \cite{Abanin15TimeDep}. 

\subsection{Complementary graphs presenting the results of the present work}

 The system phase diagram is shown in Fig. \ref{fig:PhDiagr} in  a $\delta$ - $\omega$ plane at the fixed critical transverse field $\Gamma_{c}=r\Gamma_{c0}$ ($r=1/4$) and it can be compared to the analogous phase diagram in Ref. \cite{Pollmann16PerDrive} (Fig. I there, see also Ref. \cite{Abanin16Period}). At small fields $\hbar\delta < W$  the transition frequency increases with the amplitude consistently  with Ref. \cite{Pollmann16PerDrive}. Eq. (\ref{eq:nEstSM}) results in the power law scaling $\omega \propto \delta^{\eta}$, $\eta=1+2\ln(r^{-1})/\ln(K)$, at $\hbar\delta<W/K$ changing to the direct proportionality $\omega \propto \delta$ at intermediate amplitudes $W/K<\hbar\delta<W$.  At larger fields (not considered in Refs. \cite{Pollmann16PerDrive,Abanin16Period}) the border-line frequency decreases with increasing the amplitude indicating the change in the field effect on localization at $\hbar\delta \sim W$  from field enhanced to field controlled delocalization regimes. There is no field suppressed delocalization regime since it requires $\Gamma_{c} > \Gamma_{c0}$, while here the regime $\Gamma_{c}=\Gamma_{c0}/4$ is considered. It will be interesting to verify these predictions numerically extending simulations of Ref. \cite{Pollmann16PerDrive} to larger amplitudes  and experimentally using the setup of Ref. \cite{Monthus16MultFr}. 

\begin{figure}[h!]
\centering
\includegraphics[scale=0.5]{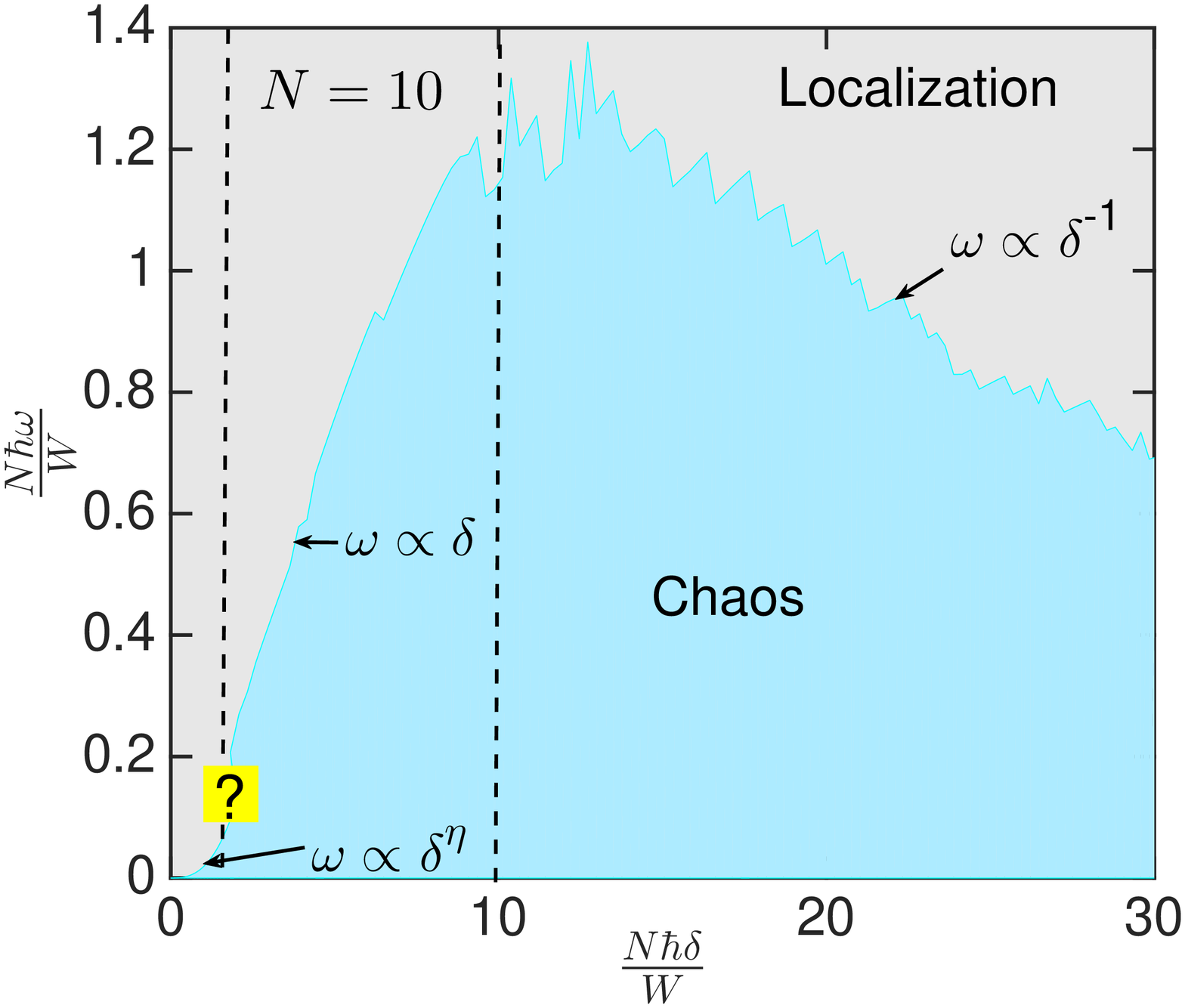}
\caption{\small Phase diagram in $\delta-\omega$ plane showing  three distinguishable behaviors determined by the numerical solution of Eq. (\ref{eq:FloqSEAbCh4}) and by asymptotic behaviors, Eqs. (\ref{eq:HiFreqFinSM}) and (\ref{eq:nEstSM}). The exponent $\eta$ determining the phase boundary scaling at small drive frequencies and amplitudes is defined as $\eta=1+2\ln(r^{-1})/\ln(K)$. Dashed lines  indicate crossovers at $\hbar\delta=W/K$ (left) and $\hbar\delta=W$ (right). The latter crossover represents the transition from field enhanced (FED) to field controlled (FCD) delocalizations. 
The question mark designates the domain $\delta \sim W/N$  that is described only qualitatively.}
\label{fig:PhDiagr}
\end{figure}

Fig. \ref{fig:GamcOm} shows the localization threshold dependence on the drive frequency. Both field enhanced and field controlled delocalization regimes are characterized by the same frequency dependence $\Gamma_{c} \propto \sqrt{\omega}$ so they can hardly be distinguished. The field suppressed regime is characterized by the strong oscillations of the localization threshold that can change by order of magnitude between maxima and minima of the Bessel function $J_{0}(\delta/\omega)$ determining the localization threshold behavior in this regime. 

\begin{figure}[h!]
\centering
\includegraphics[scale=0.5]{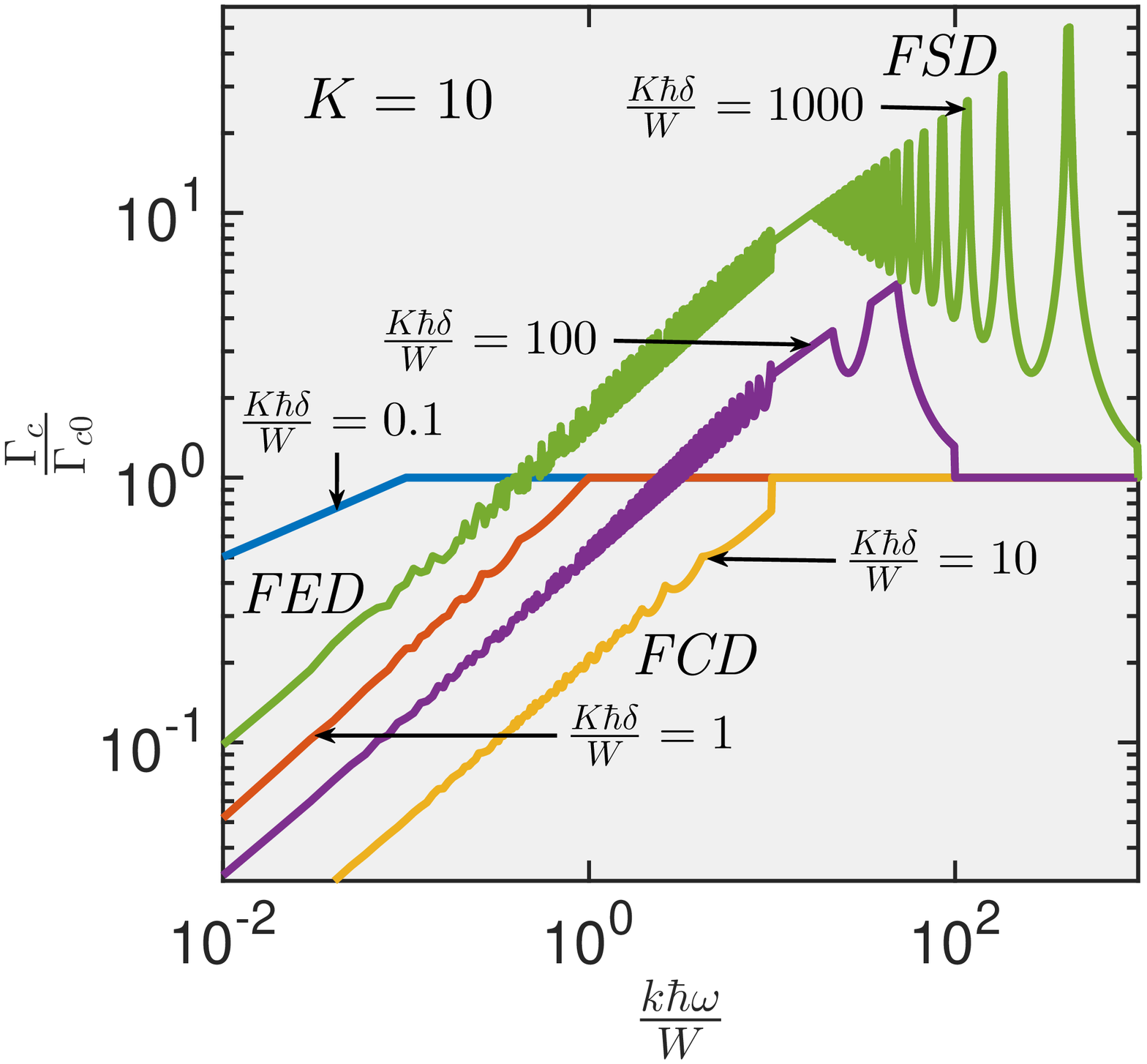}
\caption{\small Localization threshold change due to an external drive vs.  a drive frequency calculated combining the numerical solution of Eq. (\ref{eq:FloqSEAbCh4}) for $\hbar\delta >W/K$ and $\hbar\omega<W$ with asymptotic behaviors, Eqs. (\ref{eq:HiFreqFinSM}) for $\hbar\omega>\delta$ and (\ref{eq:nEstSM}) for $\hbar\delta<W/K$ or Eq. 5 within the main text. The domains of field enhanced (FED), field controlled (FCD) and field suppressed delocalizations are designated (see discussion in the text).} 
\label{fig:GamcOm}
\end{figure}

\subsection{Periodic discontinuous binary drive}

Here the localization is discussed in the case of periodic binary drive by the longitudinal field $\hbar\delta f(t)$, where $f$ is a periodic function with the period $T=2\pi/\omega$ and it is equal to $1/2$ for first halves of each $n^{th}$ period $(nT, ~(n+1/2)T)$ and $-1/2$ for second halves. The coupling strength between Floquet states different by a single spin flip, $\sigma=\pm 1$, taken within the rotation frames as in the case of a sinusoidal field can be then expressed as 
 \begin{eqnarray}
\Gamma_{n}=\Gamma A_{n}=\frac{\Gamma}{T} \int_{-T/2}^{T/2}dte^{-i|t|\delta\sigma-in\omega t}=\frac{\Gamma\delta\sigma\omega}{i\pi}\frac{1-(-1)^{n}e^{-i\frac{\sigma\pi\delta}{\omega}}}{\delta^2-n^2\omega^2}. 
\label{eq:binarySM}
\end{eqnarray} 
In the case $\delta \ll \omega$ one has $A_{n} \approx \Gamma\delta_{n0}$ and localization transition is almost insensitive to the drive. In the opposite case $\omega \ll \delta$
one can use Eq. (\ref{eq:AnsAnalit}) to determine localization thresholds in cases of intermediate and large fields, yet small frequency, $\hbar\omega \ll W$, while at a high frequency the localization threshold can be estimated generalizing the consideration of Sec. \ref{sec:LargeFreq}.  Then at a reasonably large drive amplitude $\delta > W/K$ one can obtain the localization threshold in the form
\begin{eqnarray}
\Gamma_{c} 
\approx\begin{cases}
   \frac{\pi}{2K\ln(\xi)\ln(\delta/\omega)P(0)}, ~ \frac{\Gamma_{c0}}{\hbar}<\delta < \frac{W}{\hbar},\\
    \frac{\pi\hbar\delta}{K\ln(\xi)}, ~ \omega< \frac{W}{\hbar} < \delta, ~ \xi=\frac{\hbar\delta^2}{\omega(W+\hbar\delta)},\\
\frac{2\pi\delta\Gamma_{c0}}{\omega\sqrt{1+|\sin(\pi\delta/\omega)|}}, ~ \frac{W}{\hbar} < \omega<\delta.
  \end{cases}  
\label{eq:ANSMain1SM}
\end{eqnarray}
The first, second and third lines in Eq. (\ref{eq:ANSMain1SM}) correspond to field enhanced, field controlled and field suppressed delocalization regimes similarly to the case of a sinusoidal drive, with modified analytical behaviors of $\Gamma_{c}$. 

The case $\omega < \delta < \frac{\Gamma_{c0}}{\hbar}$ is more complicated and needs a separate consideration, though preliminary estimates suggest that the localization threshold is getting reduced by the factor $\ln(\delta/\omega)^{\frac{1}{2p}}$ with $p=\frac{\ln\left(\frac{W}{\hbar\delta}\right)}{\ln(K)}$ similarly to Eq. (\ref{eq:nEstSM}).

\end{widetext}
\end{document}